\documentclass[a4paper,12pt]{article}
\usepackage[latin1]{inputenc}
\usepackage{graphicx}
\usepackage{amsmath,amssymb,euscript,array,mathrsfs}

\newcommand{\Tr}{\textnormal{Tr}}
\newcommand{\Ncal}{\mathcal{N}}
\newcommand{\EQ}[1]{\begin{equation} #1 \end{equation}}
\newcommand{\AL}[1]{\begin{subequations} \begin{align} #1
\end{align}\end{subequations}}
\newcommand{\ALlabel}[2]{\begin{subequations} \label{#2} \begin{align} #1
\end{align}\end{subequations}}
\newcommand{\SP}[1]{\begin{equation}\begin{split} #1
\end{split}\end{equation}}

\begin{document}

\title{Phase Structure of $\beta$-deformed $\Ncal = 4$ SYM on $S^3$ with Chemical Potentials}

\author{
	\textbf{Daniel Elander}
	\bigskip
	\\
	{\it Department of Physics}, \\ {\it Swansea University}, \\ {\it Swansea, SA2 8PP, UK}
	\bigskip
	\\
	{\tt pyde@swansea.ac.uk}
}

\date{}

\maketitle

\abstract{We study the $\beta$-deformation of $\Ncal = 4$ SYM on $S^3$ with chemical potentials for the $U(1)_R$ as well as the two global $U(1)$ symmetries. The one-loop effective potential at weak coupling is computed for both the Coulomb and Higgs branches. At near critical chemical potential and small finite temperature, we find a metastable state at the origin of moduli space. On the Higgs branch, this has the interpretation in terms of deconstruction as an extra-dimensional torus which becomes metastable for infinite size and decays to zero size through quantum tunnelling and thermal activation. At strong coupling, the theory is described by its gravitational dual. The relevant background is found by performing a TsT-transformation on the solution describing an $AdS_5$ black hole spinning in $S^5$. A probe-brane calculation, using giant gravitons as probes, reveals qualitative agreement with the weak coupling results.}

\pagebreak
\tableofcontents
\pagebreak

\section{Introduction}
One of the many exciting results to have come out of the $AdS/CFT$ correspondence \cite{Maldacena:1997re, Witten:1998qj, Gubser:1998bc} is that $\Ncal=4$ supersymmetric Yang-Mills at finite temperature is related to black holes in $AdS_5$. For example, the Hawking-Page phase transition \cite{Hawking:1982dh}, in which a black hole forms above a critical value of the temperature, turns out to be dual, by the correspondence, to a confinement-deconfinement phase transition in the quantum field theory on the boundary \cite{Witten:1998zw}. The link between the thermodynamics of black holes and that of $\Ncal=4$ SYM makes it an interesting project to map out the phase structure, and compare the results at strong and weak 't~Hooft coupling. Much effort has been devoted towards this goal \cite{Sundborg:1999ue, Aharony:2003sx, Aharony:2005bq, Liu:2004vy, Yamada:2006rx, Harmark:2006di, Hollowood:2006xb, Harmark:2006ta, Harmark:2006ie, Harmark:2007px, AlvarezGaume:2005fv, Basu:2005pj, AlvarezGaume:2006jg, Azuma:2007fj, Dutta:2007ws, Yamada:2008em, Hollowood:2008gp, Murata:2008bg, Hawking:1999dp, Cvetic:1999ne, Yamada:2007gb, Hawking:1998kw}.

In this article, we study a marginal deformation of $\Ncal=4$ SYM known as the $\beta$-deformation, which changes the superpotential of $\Ncal=4$ SYM to
\EQ{
	W = i 2 \sqrt{2} \Tr \left( e^{i\pi\beta} \Phi_1 \Phi_2 \Phi_3 - e^{-i\pi\beta} \Phi_1 \Phi_3 \Phi_2 \right),
}
where $\beta$ is the deformation parameter. While the $\beta$-deformation breaks the amount of supersymmetry to $\Ncal=1$, it is interesting in that it preserves the conformal invariance of the original theory \cite{Leigh:1995ep}. The $SO(6)$ R-symmetry of $\Ncal=4$ SYM is broken to $U(1)^3$, and we can add chemical potentials $\mu_i$ for these three $U(1)$s. The addition of chemical potentials breaks the conformal invariance, as well as all the supersymmetry of the theory. Furthermore, a negative mass squared term $-\mu_i^2$ gets generated for the scalars charged under the associated symmetry, and therefore the theory becomes unstable, unless it is defined at finite volume where the scalars also couple to the curvature through the conformal coupling, thus generating positive mass squared terms. We will define the $\beta$-deformed theory on $S^1 \times S^3$, where $S^1$ is the compactified time direction. In particular, we will be interested in chemical potentials which are close to critical, meaning that the negative mass squared terms that they generate almost cancel the ones from the conformal coupling. Classically, it is only for critical chemical potentials that there are flat directions and a non-trivial moduli space. The moduli space of the $\beta$-deformed theory has a Coulomb branch, and also, for special values of the deformation parameter $\beta$, additional Higgs branches open up \cite{Berenstein:2000ux, Dorey:2004xm, Berenstein:2000hy}. On these branches the theory is equivalent at low energies to $\Ncal = 4$ SYM. At intermediate energies, it can be viewed as the deconstruction of $\Ncal = (1,1)$ SYM in six dimensions, with the two extra dimensions forming a latticized torus \cite{Dorey:2003pp, Dorey:2004iq}. In essence, the torus forms because we can reinterpret the two gauge group indices of the adjoint scalars as discretized extra dimensions.

At finite temperature, the gravitational dual of $\Ncal=4$ SYM with chemical potentials and gauge group $SU(N)$ is a solution of $\Ncal=2$ five-dimensional $U(1)^3$ gauged supergravity, which describes a Reissner-Nordström black hole carrying charges with respect to the three $U(1)$s \cite{Behrndt:1998ns, Behrndt:1998jd}. There are three background gauge fields $A_{i\mu}^{(1)}$ whose values at the boundary correspond to the value of the chemical potentials of the boundary quantum field theory. The five-dimensional charged black hole solution can be embedded in ten dimensional Type IIB supergravity compactified on $S^5$ \cite{Cvetic:1999xp}. The resulting Type IIB supergravity solution describes an (uncharged) $AdS_5$ black hole rotating in $S^5$. In \cite{Lunin:2005jy}, it was described how to, in general, generate the ten-dimensional solution describing the $\beta$-deformed theory by performing a TsT-transformation, a T-duality followed by a shift of variables and then another T-duality, on the solution describing $\Ncal=4$ SYM. Applying this method to the ten-dimensional rotating black hole solution, we obtain the Type IIB supergravity solution which is the gravity dual of finite temperature $\beta$-deformed $\Ncal=4$ SYM with chemical potentials.

It was found in \cite{Hollowood:2008gp} that, at zero temperature, $\Ncal=4$ SYM on $S^1 \times S^3$ with critical chemical potentials has a one-loop effective action that is independent of the scalar VEVs. In this article, we repeat the calculation for the $\beta$-deformed theory and find the same result for $SU(N)$ gauge group, but a different one for $U(N)$. Since, for gauge group $U(N)$, the overall $U(1)$ decouples for $\Ncal=4$ SYM, this could not have happened in that case. However, in the $\beta$-deformed theory, it is no longer true that the overall $U(1)$ decouples. At finite temperature, and near critical chemical potential, $\Ncal=4$ SYM has a metastable state at the origin of moduli space, which decays through thermal activation or quantum tunnelling due to the runaway behaviour of the potential for large values of the scalar VEVs \cite{Hollowood:2008gp}. This is also true for the Coulomb branch of the $\beta$-deformed theory. We perform calculations which show that the same is true for the Higgs branch, where an interpretation can be made in which the extra-dimensional torus has a metastable state when its volume is infinite, that then decays to zero volume.

In order to see if the picture remains qualitatively the same at strong 't~Hooft coupling, we perform a probe-brane calculation in the dual gravitational background. This was done for finite temperature $\Ncal=4$ SYM in \cite{Yamada:2008em}, where it was found that for near critical chemical potentials, there is a metastable phase at strong coupling. On the Coulomb branch, the probe-branes we will use are D3-brane giant gravitons, which extend along the three non-radial spatial coordinates of $AdS_5$, whereas on the Higgs branch, the probe-branes are D5-brane giant gravitons, which, in addition to extending along the same three coordinates in $AdS_5$ as the D3-branes, also wrap around the torus formed by the two directions in $S^5$ which involve the TsT-transformation. Giant gravitons were studied in, for example, \cite{McGreevy:2000cw, Grisaru:2000zn, Hashimoto:2000zp, Gubser:1998jb}, and in \cite{Imeroni:2006rb, Pirrone:2006iq} they were studied in the Lunin-Maldacena background. We show that for near critical chemical potentials, the metastable phases of $\beta$-deformed $\Ncal=4$ SYM at finite temperature persist at strong 't~Hooft coupling.

The structure of this article is as follows. In section~2, we review how to add chemical potentials to the theory, and the moduli space of $\beta$-deformed $\Ncal=4$ SYM. In section~3, we compute the one-loop effective action for the theory on the Coulomb branch, whereas in section~4 we do the same for the Higgs branch. Section~5 covers the metastable phases that occur at finite temperature and near critical chemical potentials. In section~6, we find the gravity dual describing the $\beta$-deformed theory, and in section~7, we carry out the probe-brane calculation which establishes the existence of a metastable phase at strong 't~Hooft coupling. Finally, we summarize our results in section~8.

\section{The $\beta$-deformation of $\mathcal{N}=4$ SYM}
\subsection{Chemical Potentials}
The $\beta$-deformation of $\mathcal{N}=4$ SYM is obtained by deforming the $\mathcal{N}=4$ superpotential to (our conventions are explained in the Appendix)
\EQ{
	W = i 2 \sqrt{2} \Tr \, \Phi_1 [\Phi_2,\Phi_3]_\beta,
}
where
\EQ{
	[A,B]_\beta \equiv e^{i\pi\beta} AB - e^{-i\pi\beta} BA,
}
and $\mathcal{N}=4$ SYM corresponds to $\beta=0$. In the following, we shall take $\beta$ to be real.

A non-zero $\beta$ breaks the original $SU(4)$ R-symmetry to $U(1)^3$, where each of the $\Phi_i$ is charged under a
different $U(1)$. For the complex scalars $\phi_i$, we write this as
\SP{
	\hat Q_1 (\phi_1,\phi_2,\phi_3) &= (1,0,0), \\
	\hat Q_2 (\phi_1,\phi_2,\phi_3) &= (0,1,0), \\
	\hat Q_3 (\phi_1,\phi_2,\phi_3) &= (0,0,1),
}
and similarly for the fermions:
\SP{
	\hat Q_1 (\lambda,\chi_1,\chi_2,\chi_3) &= \tfrac{1}{2} (1,1,-1,-1), \\
	\hat Q_2 (\lambda,\chi_1,\chi_2,\chi_3) &= \tfrac{1}{2} (1,-1,1,-1), \\
	\hat Q_3 (\lambda,\chi_1,\chi_2,\chi_3) &= \tfrac{1}{2} (1,-1,-1,1).
}

The grand canonical partition function is
\EQ{
	Z(T,\mu_i) = \Tr \, e^{-\frac{1}{T} (\hat H - \sum_i \mu_i \hat Q_i ) },
}
where $\mu_i$ are the chemical potentials. Viewed as a Euclidean path integral with time compactified on $S^1$, adding chemical potentials to the theory is equivalent to letting \cite{Yamada:2006rx}
\EQ{
	D_\mu \rightarrow D_\mu - \delta_{\mu,0} \sum_i \mu_i \hat Q_i.
}
Hence, the kinetic terms for the complex scalars have the form
\EQ{
	\label{eq:kineticscalars}
	2 \Tr \, \left(D_\mu + \mu_i \delta_{\mu,0} \phi_i \right)^\dagger (D_\mu - \mu_i \delta_{0,\mu}) \phi_i,
}
whereas for the fermions the kinetic terms are
\EQ{
	2 \Tr \, \bar\chi_i ( i \sigma_\mu D^\mu - i \bar\mu_i ) \chi_i,
}
where
\SP{
	\bar\mu_0 &= \tfrac{1}{2} (\mu_1 + \mu_2 + \mu_3), \\
	\bar\mu_1 &= \tfrac{1}{2} (\mu_1 - \mu_2 - \mu_3), \\
	\bar\mu_2 &= \tfrac{1}{2} (-\mu_1 + \mu_2 - \mu_3), \\
	\bar\mu_3 &= \tfrac{1}{2} (-\mu_1 - \mu_2 + \mu_3),
}
and we have made the definition $\chi_0 \equiv \lambda$.

We note that we can also use a basis with one $U(1)_R$ and two global $U(1)$, where the global $U(1)$s are linear combinations of the original three $U(1)_R$s. It is therefore not unreasonable to expect that the results will be qualitatively different when one turns on two of the chemical potentials $\mu_i$ from when one only turns on one.

\subsection{Classical Moduli Space}
With the theory defined on $S^3$, there is a conformal coupling of the scalars to the curvature which takes the form
\EQ{
	2 \Tr \, R^{-2} \phi_i^\dagger \phi_i.
}
Furthermore, from \eqref{eq:kineticscalars} we get a similar term but with opposite sign:
\EQ{
	- 2 \Tr \, \mu_i^2 \phi_i^\dagger \phi_i.
}
Only when at least one of the $\mu_i$ has the critical value $\mu_i = R^{-1}$ is there any possibility of flat directions and a non-trivial moduli space. However, the F- and D-flatness conditions also need to be satisfied:
\EQ{
	[\phi_1,\phi_2]_\beta = [\phi_2,\phi_3]_\beta = [\phi_3,\phi_1]_\beta = 0,
}
\EQ{
	\sum_{i=1}^3 [\phi_i^\dagger,\phi_i] = 0.
}
These are solved by giving each $\phi_i$ a diagonal VEV, while imposing the restriction that for each row (equivalently column), no more than one of $\phi_i$ is allowed to have a non-zero entry. We also have to mod out by the Weyl group. This defines the Coulomb branch, where, for generic VEVs, the original $SU(N)$ gauge symmetry is broken down to $U(1)^{N-1}$, and the $U(N)$ gauge symmetry is broken to $U(1)^N$.

For rational values of $\beta$, there are additional Higgs branches. For example, we can take $\beta = 1/N$ and give VEVs to the scalars as
\AL{
	\left\langle \phi_1 \right\rangle &= \lambda^{(1)} U_{(N)}, \\
	\left\langle \phi_2 \right\rangle &= \lambda^{(2)} V_{(N)}, \\
	\left\langle \phi_3 \right\rangle &= \lambda^{(3)} V_{(N)}^\dagger U_{(N)}^\dagger,
}
where $\lambda^{(1)}$, $\lambda^{(2)}$, and $\lambda^{(3)}$ are complex numbers, and
\AL{
	U_{(N)} &= \textnormal{diag} \left( \omega, \omega^2, \ldots, \omega^N \right) \\
	\left(V_{(N)}\right)_{ab} &=
	\begin{cases}
	1 \textnormal{	if } b=a+1 \textnormal{ mod } N, \\
	0 \textnormal{ otherwise}
	\end{cases},
}
with $\omega = e^{2\pi i \beta}$. This breaks $U(N)$ to $U(1)$, while in the $SU(N)$ case the gauge group is completely broken. To obtain the right moduli space, we also have to mod out by the discrete gauge transformations
\EQ{
	\phi_i \rightarrow \Gamma_j \phi_i \Gamma_j^{\dagger},
}
where $\Gamma_1 = U_{(N)}$ and $\Gamma_2 = V_{(N)}$. These rotate $\lambda^{(i)}$ by discrete phases $\omega$ \cite{Dorey:2003pp}. After taking this to account, the moduli space is ${\mathbb C}^3/({\mathbb Z}^N \times {\mathbb Z}^N)$.

More generally, we have the solution
\AL{
	\left\langle \phi_1 \right\rangle &= \Lambda^{(1)} \otimes U_{(n)}, \\
	\left\langle \phi_2 \right\rangle &= \Lambda^{(2)} \otimes V_{(n)}, \\
	\left\langle \phi_3 \right\rangle &= \Lambda^{(3)} \otimes V_{(n)}^\dagger U_{(n)}^\dagger,
}
with
\EQ{
	\Lambda^{(i)} = \textnormal{diag} \left( \lambda^{(i)}_1, \lambda^{(i)}_2, \ldots, \lambda^{(i)}_m \right)
}
and $N=nm$, $\beta=1/n$. For generic $\Lambda^{(i)}$, this breaks $U(N)$ to $U(1)^m$, and $SU(N)$ to $U(1)^{m-1}$. The low energy theory turns out to be $\mathcal{N}=4$ on the Coulomb branch \cite{Dorey:2003pp, Dorey:2004iq}.

\section{One-Loop Effective Potential for the Coulomb Branch}
\subsection{General Considerations}
We will now compute the Wilsonian one-loop effective potential by integrating out all but the lightest fields of the theory. The fields can be expanded on $S^3 \times S^1$ in terms of spherical harmonics and Matsubara modes. The analysis is similar to that in \cite{Hollowood:2008gp}. We will turn on one critical chemical potential $\mu_1 = R^{-1}$ and give a background VEV to the mode constant on $S^3 \times S^1$ of the associated complex scalar
\EQ{
	\phi_1 \rightarrow \frac{\varphi}{\sqrt{2}} + \phi_1.
}
In addition, there will be a background value for the spatial zero mode of the holonomy of the time component of the gauge field around the thermal circle:
\EQ{
	A_0 \rightarrow \alpha + A_0.
}
The effective action is parametrized by $\varphi$, and $\alpha$, which in this section we shall take to both be diagonal.

The Lagrangian for the bosons and the ghosts $(\bar c,c)$ at second order is (see Appendix for details)
\SP{
	\mathcal{L}_b^{(2)} = \frac{1}{g^2} 2 \Tr \bigg( \frac{1}{2} A_0 ( - \tilde{D}_0^2 - \Delta^{(s)} + \varphi^\dagger \varphi ) A_0 + \\
	+ \frac{1}{2} A_i ( - \tilde{D}_0^2 - \Delta^{(v)} + \varphi^\dagger \varphi ) A_i + \\
	+ \bar{c} ( - \tilde{D}_0^2 - \Delta^{(s)} + \varphi^\dagger \varphi ) c + \\
	+ \phi_1^\dagger (-\tilde{D}_0^2 - \Delta^{(s)} + \varphi^\dagger \varphi ) \phi_1 + \\
	+ \mu_1 \left[ 2 \phi_1^\dagger \tilde{D}_0 \phi_1 + i \sqrt{2} \phi_1^\dagger \varphi A_0 + i \sqrt{2} A_0 \varphi^\dagger \phi_1 \right] + \\
	+ \phi_2^\dagger ( - \tilde{D}_0^2 - \Delta^{(s)} + \varphi_\beta^\dagger \varphi_\beta + R^{-2} ) \phi_2 + \\
	+ \phi_3^\dagger ( - \tilde{D}_0^2 - \Delta^{(s)} + \varphi_{-\beta}^\dagger \varphi_{-\beta} + R^{-2} ) \phi_3
	\bigg),
}
where we have used the notation
\AL{
	\varphi &\equiv [\varphi, \cdot] \\
	\varphi_\beta &\equiv [\varphi, \cdot]_\beta \\
	\varphi_\beta^\dagger &\equiv [\varphi^\dagger, \cdot]_{-\beta},
}
and fixed the gauge by adding a term
\EQ{
	\mathcal{L}_{gf} = \frac{1}{g^2} \Tr \left( \nabla_i A^i + \tilde{D}_0 A^0 - \frac{i}{\sqrt{2}} \left( \varphi^\dagger \phi_1 + \varphi \phi_1^\dagger \right) \right)^2
}
to the Lagrangian, corresponding to $R_\xi$-gauge with Feynman parameter $\xi=1$. $\Delta^{(s)}$ and $\Delta^{(v)}$ are the scalar and vector Laplacians on $S^3$ respectively.

For the fermions, we have
\SP{
		\mathcal{L}_f^{(2)} = 2 \Tr \bigg( \sum_{i=0}^3 \bar{\chi_i} ( i \sigma_\mu D^\mu - i \bar{\mu_i} ) \chi_i -
	\chi_0 (i \varphi^\dagger) \chi_1 - \bar{\chi}_1 ( -i \varphi) \bar{\chi}_0 - \\
	- \chi_3 (i \varphi_\beta) \chi_2 - \bar{\chi}_2 ( -i \varphi_\beta^{\dagger}) \bar{\chi}_3 \bigg)
}

The one-loop correction to the effective potential is given by
\EQ{
	\label{eq:logdeteff}
	V_1 = \frac{T}{2\pi^2R^3} \frac{1}{2} \sum_{\textnormal{species}} \sum_{ij}^N \sum_{\ell = \ell_0}^\infty d_\ell^{B(F)} \log \det \left( - \tilde D_0^2 + \varepsilon_\ell \left( \varphi \right)  \right),
}
where $\ell$ is the angular momentum quantum number of the mode with $\ell_0$ its lowest value, $d_\ell^{B(F)}$ is the degeneracy including differing signs for bosons and fermions, and finally $\varepsilon_\ell$ is the energy of the mode. Their values for gauge group $U(N)$ are summarized in Table 1. All possible sign combinations are allowed. Note also that the expressions are only valid at vanishing or critical chemical potentials.\footnote{More precisely, the expressions for the fermions are valid for any values of the chemical potentials, but the chemical potentials need to be vanishing or critical in order for the expressions for the complex scalars and $A_0$ to take the simple form of Table 1.} Since $(\varphi \phi)_{ij} = (\varphi_i - \varphi_j ) \phi_{ij}$, it makes sense to define
\EQ{
	\varphi_{ij} \equiv \varphi_i - \varphi_j,
}
where $\varphi_i$ is the $i$th diagonal component of $\varphi$,
and similarly for the beta-commutator
\EQ{
	{\varphi_\beta}_{ij} \equiv e^{i\pi\beta} \varphi_i - e^{-i\pi\beta} \varphi_j.
}
$\varepsilon_\ell(\varphi)$ should then be thought of as a function of $\varphi_{ij}$ and ${\varphi_\beta}_{ij}$.

\begin{table}
\begin{center}
\begin{tabular}{c | c | c | c}
  Field & $d_\ell$ & $|\varepsilon_\ell|$ & $\ell_0$ \\ \hline
 	$B_i$ & $2\ell(\ell+2)$ & $\sqrt{R^{-2}(\ell+1)^2 + \varphi^\dagger \varphi}$ & 1 \\
 	$C_i$ & $(\ell+1)^2$ & $\sqrt{R^{-2}\ell(\ell+2) + \varphi^\dagger \varphi}$ & 1 \\
 	$(c,\bar{c})$ & $-2(\ell+1)^2$ & $\sqrt{R^{-2}\ell(\ell+2) + \varphi^\dagger \varphi}$ & 0 \\
 	$(A_0,\phi_1,\phi_1^\dagger)_1$ & $(\ell+1)^2$ & $\sqrt{R^{-2}\ell(\ell+2) + \varphi^\dagger \varphi}$ & 0 \\
 	$(A_0,\phi_1,\phi_1^\dagger)_{2,3}$ & $(\ell+1)^2$ & $\sqrt{R^{-2}(\ell+1 \pm R \mu_1)^2 + \varphi^\dagger \varphi}$ & 0 \\
 	$\phi_2$ & $(\ell+1)^2$ & $\sqrt{R^{-2}(\ell+1)^2 + \varphi_\beta^\dagger \varphi_\beta} \pm \mu_2$ & 0 \\
 	$\phi_3$ & $(\ell+1)^2$ & $\sqrt{R^{-2}(\ell+1)^2 + \varphi_{-\beta}^\dagger \varphi_{-\beta}} \pm \mu_3$ & 0 \\
 	$(\lambda,\chi_1)$ & $-\ell(\ell+1)$ & $	\sqrt{  R^{-2} \left( \ell + \frac{1}{2} \pm \frac{R \mu_1}{2} \right)^2 + \varphi^\dagger \varphi} \pm \frac{\mu_2 + \mu_3}{2}$ & 1 \\
 	$(\chi_2,\chi_3)$ & $-\ell(\ell+1)$ & $	\sqrt{  R^{-2} \left( \ell + \frac{1}{2} \pm \frac{R \mu_1}{2} \right)^2 + \varphi_\beta^\dagger \varphi_\beta} \pm \frac{\mu_2 - \mu_3}{2}$ & 1
\end{tabular}
\end{center}
\caption{\footnotesize The energies $\varepsilon_\ell$ for the Coulomb branch and gauge group $U(N)$, together with their degeneracies for the various fields. The expressions are valid for vanishing or critical ($\mu_i = R^{-1}$) chemical potentials. All possible sign combinations are allowed.}
\end{table}

After a Poisson resummation over the Matsubara frequencies, \eqref{eq:logdeteff} can be recast as a sum over species \cite{Hollowood:2008gp} with bosons contributing
\EQ{
	\label{eq:bpoissonsum}
	\frac{1}{\textnormal{Vol}(S^3)} \frac{1}{2} \sum_{i,j=1}^N \sum_{\ell = \ell_0}^\infty d_\ell^B \left( |\varepsilon_\ell(\varphi)| - T \sum_{k=1}^\infty \frac{1}{k} e^{-\frac{k}{T} |\varepsilon_\ell(\varphi)|} \cos(k \alpha_{ij} / T) \right),
}
and fermions contributing
\EQ{
	\label{eq:fpoissonsum}
	\frac{1}{\textnormal{Vol}(S^3)} \frac{1}{2} \sum_{i,j=1}^N \sum_{\ell = \ell_0}^\infty d_\ell^F \left( |\varepsilon_\ell(\varphi)| - T \sum_{k=1}^\infty \frac{(-1)^k }{k} e^{-\frac{k}{T} |\varepsilon_\ell(\varphi)|} \cos(k \alpha_{ij} / T) \right),
}
where $\alpha_{ij} = \alpha_i - \alpha_j$ ($\alpha_i$ refers to the $i$th diagonal component of $\alpha$).

\subsection{Zero Temperature}
At zero temperature, only the Casimir energy parts of \eqref{eq:bpoissonsum} and \eqref{eq:fpoissonsum} contribute:
\EQ{
	V_1(T=0) = \frac{1}{\textnormal{Vol} (S^3)} \frac{1}{2} \sum_{\textnormal{species}} \sum_{i,j=1}^N \sum_{\ell = \ell_0}^\infty d_\ell^{B(F)} |\varepsilon_\ell(\varphi)|.
}
We regularize this expression by introducing a cut-off that does not depend on the chemical potentials, as follows \cite{Hollowood:2008gp}:
\EQ{
	\frac{1}{2} \sum_{i,j=1}^N \sum_{l=l_0}^\infty d_l^{B(F)} |\epsilon_\ell(\varphi)| f(\left|\epsilon_\ell(\varphi)|_{\mu_i=0}\right|/\Lambda),
}
where $\Lambda$ is the cut-off, and $f(x)$ is a function that is equal to 1 for $x \leq 1$ and zero for $x>1$.

Since
\EQ{
	(\varphi_\beta^\dagger \varphi_\beta \phi)_{ij} = |e^{i\pi\beta} \varphi_i - e^{-i\pi\beta} \varphi_j|^2 \phi_{ij},
}
\EQ{
	(\varphi_{-\beta}^\dagger \varphi_{-\beta} \phi)_{ij} = |e^{i\pi\beta} \varphi_j - e^{-i\pi\beta} \varphi_i|^2 \phi_{ij},
}
and we sum over $i$ and $j$, there is no need to distinguish between the two in the calculation. Furthermore, we note that $C_i$, $(c,\bar{c})$, $(A_0,\phi_1,\phi_1^\dagger)_1$, and the contribution given by $(A_0,\phi_1,\phi_1^\dagger)_{2,3}$ with a minus sign and $\ell=0$ cancel against each other. Converting the sum over $\ell$ into an integral for the remaining fields by using the Abel-Plana formula \cite{Mostepanenko:1997sw}
\EQ{
	\sum_{n=0}^\infty F(n) = \int_0^\infty dx \, F(x) + \frac{1}{2} F(0) - 2 \int_0^\infty dx \frac{\textnormal{Im} F(ix)}{e^{2\pi x} - 1},
}
and summing over the species, we find the zero temperature effective potential
\SP{
	\label{eq:zeroTeffpot}
	\textnormal{Vol}(S^3) \, V_1(T=0) = \frac{3N^2}{16 R} + \frac{R}{8} \Tr  \left( \varphi_\beta^\dagger \varphi_\beta - \varphi^\dagger \varphi \right) = \\
	= \frac{3N^2}{16 R} + \frac{R}{8} \sum_{i,j=1}^N \left( |e^{i\pi\beta} \varphi_i - e^{-i\pi\beta} \varphi_j|^2 - |\varphi_i - \varphi_j|^2 \right) = \\
	= \frac{3N^2}{16R} + \frac{R}{2} \sin^2(\pi \beta) \left| \sum_{i=1}^N \varphi_i \right| ^2.
}
Although we have used the expressions of the energies for gauge group $U(N)$, in the large $N$ limit this expression is valid for $SU(N)$ also. The reason is that the only energies which are affected in going from $U(N)$ to $SU(N)$ are those for the diagonal fluctuations. For gauge group $SU(N)$, \eqref{eq:zeroTeffpot} reduces to the same expression as in the $\mathcal{N}=4$ case. For $U(N)$, the result is sensitive to the overall $U(1)$ which, unlike in the $\mathcal{N}=4$ theory, does not decouple from the dynamics for generic $\beta$. We note that \eqref{eq:zeroTeffpot} also is valid when $\mu_2$ or $\mu_3$ are critical, since they appear outside the square roots with plus or minus signs in the expressions for the energies and therefore cancel against each other when we sum all modes. Therefore, at zero temperature there is no difference between turning on a chemical potential for a $U(1)_R$ or a global $U(1)$. (We note that even though there is a positive mass squared for the traceful part of $\phi$, there is no metastable phase for near (and above) critical chemical potential due to the fact that the traceless modes still have negative masses squared.)

We also note briefly that using another Abel-Plana formula \cite{Mostepanenko:1997sw}
\EQ{
	\sum_{n=0}^\infty F(n+1/2) = \int_0^\infty dx \, F(x) + 2 \int_0^\infty dx \frac{\textnormal{Im} F(ix)}{e^{2\pi x} + 1},
}
we can derive an expression for the off-shell effective action without chemical potentials:
\SP{
	\label{eq:offshelleffaction}
	\textnormal{Vol}(S^3) V = \frac{R}{8} \left( |\varphi|^2 - |\varphi_\beta|^2 \right) + \\ +
	\frac{R}{4} \left( |\varphi_\beta|^2 - |\varphi|^2 \right) \log \left( 2 R \Lambda \right) + \\ +
	\frac{R}{4} |\varphi|^2 \log \left( R |\varphi| \right) - \frac{R}{4} |\varphi_\beta|^2 \log \left( R |\varphi_\beta| \right) + \\ +
	R^{-1} \int_{R|\varphi|}^\infty dl \, \frac{\left(4l^2 + \frac{3}{2} + \frac{1}{2} e^{-2\pi l} \right) \sqrt{l^2 - R^2 |\varphi|^2}}{\sinh(2\pi l)} + \\
	+ R^{-1} \int_{R|\varphi_\beta|}^\infty dl \, \frac{\left(4l^2 + \frac{1}{2} - \frac{1}{2} e^{-2\pi l} \right) \sqrt{l^2 - R^2 |\varphi_\beta|^2}}{\sinh(2\pi l)}.
}
In the above expression, the sum over $i$ and $j$ is implicit. Since
\EQ{
 \sum_{ij} \left( |\varphi_\beta|^2 - |\varphi|^2 \right) = 4 \sin^2 \left( \pi \beta \right) \left| \sum_{i} \varphi_i \right|^2,
}
there is no ultraviolet divergence in the $SU(N)$ theory. Note also, that the ultraviolet divergence for the $U(N)$ theory is a finite volume effect. For $\beta = 0$ and $\mathcal{N}=4$ SYM, \eqref{eq:offshelleffaction} reduces to
\EQ{
	2 R^{-1} \int_{R|\varphi|}^\infty dl \, \frac{\left(4l^2 + 1 \right) \sqrt{l^2 - R^2 |\varphi|^2}}{\sinh(2\pi l)},
}
first computed in \cite{Hollowood:2006xb}.

\subsection{Finite Temperature}
Consider
\EQ{
		\sum_{l = l_0}^\infty d_l^{B(F)} e^{-\frac{k}{T} |\epsilon_l(\varphi)|}
}
appearing in the temperature dependent part of the expression \eqref{eq:bpoissonsum} and \eqref{eq:fpoissonsum} for the bosonic and fermionic contributions to the one-loop effective potential. Summing over all bosonic modes, we obtain
\SP{
	\sum_{l = 1}^\infty 4l^2 \bigg( e^{-\frac{k}{T} \sqrt{R^{-2} l^2 + |\varphi|^2}} + \\ +
	\frac{1}{2} \left[ \cosh \left( \frac{k \mu_2}{T} \right) + \cosh \left( \frac{k \mu_3}{T} \right) \right] e^{-\frac{k}{T} \sqrt{R^{-2} l^2 + |\varphi_\beta|^2}}  \bigg).
}
Similarly for the fermionic modes, we have
\SP{
	- \sum_{l = 1}^\infty 4l^2 \bigg(
	\cosh \left( \frac{k (\mu_2 +\mu_3)}{2T} \right)	e^{-\frac{k}{T} \sqrt{R^{-2} l^2 + |\varphi|^2}} + \\ + \cosh \left( \frac{k (\mu_2 - \mu_3)}{2T} \right) e^{-\frac{k}{T} \sqrt{R^{-2} l^2 + |\varphi_\beta|^2}}  \bigg)
}
Hence, the full expression for the one-loop effective potential at finite temperature is
\SP{
	\label{eq:finitetempeffact}
	V_0 + V_1 = \frac{1}{\textnormal{Vol}(S^3)} \bigg\{
	\frac{3N^2}{16R} + \frac{R}{2} \sin^2(\pi \beta) \left| \sum_{i=1}^N \varphi_i \right| ^2  - \\ -
	2T \sum_{i,j=1}^N
		\sum_{k=1}^\infty \frac{\cos(k \alpha_{ij} / T)}{k}
		\sum_{l = 1}^\infty l^2 \bigg( \left[ 1 - (-1)^k \cosh \left( \frac{k (\mu_2 +\mu_3)}{2T} \right) \right] \\ e^{-\frac{k}{T} \sqrt{R^{-2} l^2 + |\varphi_{ij}|^2}} +
	\frac{1}{2} \bigg[ \cosh \left( \frac{k \mu_2}{T} \right) + \cosh \left( \frac{k \mu_3}{T} \right) - \\ - 2(-1)^k \cosh \left( \frac{k (\mu_2 - \mu_3)}{2T} \right) \bigg] e^{-\frac{k}{T} \sqrt{R^{-2} l^2 + |{\varphi_\beta}_{ij}|^2}}  \bigg)
	\bigg\}
}
Since there is an attractive potential for the $\alpha_i$, we can put $\alpha_{ij}=0$, which means that the theory is in the deconfined phase. We see that unlike in the zero temperature case, because $\varphi_\beta$ appears in the exponential, there is now a non-trivial dependence on $\beta$ not just for the overall $U(1)$, but also for $SU(N)$.

\section{One-Loop Effective Potential for the Higgs Branch}
Let us first work out the case $n=N$, $\beta = 1/N$. To simplify matters, we will only give VEVs to two of the complex scalars
\AL{
	\phi_1 \rightarrow \frac{\lambda^{(1)} U_{(N)}}{\sqrt{2}} + \phi_1, \\
	\phi_2 \rightarrow \frac{\lambda^{(2)} V_{(N)}}{\sqrt{2}} + \phi_2.
}
In order to be able to do this, we need to (at least) turn on the two chemical potentials $\mu_1 = \mu_2 = R^{-1}$. Although technically more involved, conceptually the calculation of the one-loop effective potential for the Higgs branch proceeds in the same way as that for the Coulomb branch. The details are outlined in the Appendix. Crucially, one expands the fluctuations as
\EQ{
	\phi = \frac{1}{\sqrt{2N}} \sum_{i,j=1}^N \phi_{i,j} J_{i,j},
}
where
\EQ{
	J_{a,b} \equiv V_{(N)}^a U_{(N)}^{-b} \omega^{\frac{ab}{2}}
}
is a basis for $N \times N$ matrices \cite{Dorey:2003pp}. Here, $a$ and $b$ are integers defined modulo $N$. Using that
\AL{
	U_{(N)} &= J_{0,-1} \\
	V_{(N)} &= J_{1,0},
}
and the commutation relations
\AL{
	[J_{a,b},J_{c,d}] &= 2 \sin \left( \frac{(bc-ad)\pi}{N} \right) J_{a+c,b+d}, \\
	[J_{a,b},J_{c,d}]_{\pm \beta} &= 2 \sin \left( \frac{(bc-ad \pm 1)\pi}{N} \right) J_{a+c,b+d},
}
it is possible to derive the energies for gauge group $U(N)$ summarized in Table 2. The form of the spectrum has the interpretation as the appearance of two extra compact dimensions forming a discretized torus \cite{Dorey:2003pp} with radii given by
\ALlabel{
	R_1 &= \frac{N}{2\pi|\lambda^{(1)}|}, \\
	R_2 &= \frac{N}{2\pi|\lambda^{(2)}|},
}{eq:torusradii}
and lattice spacings given by
\AL{
	\epsilon_1 &= \frac{2 \pi R_1}{N} = \frac{1}{|\lambda^{(1)}|}, \\
	\epsilon_2 &= \frac{2 \pi R_2}{N} = \frac{1}{|\lambda^{(2)}|}.
}

\begin{table}
\begin{center}
\begin{tabular}{c | c | c | c}
  Field & $d_\ell$ & $|\varepsilon_\ell|$ & $\ell_0$ \\ \hline
 	$B_i$ & $2\ell(\ell+2)$ & $\sqrt{R^{-2}(\ell+1)^2 + X_{ij}(\lambda^{(1,2)})}$ & 1 \\
 	$C_i$ & $(\ell+1)^2$ & $\sqrt{R^{-2}\ell(\ell+2) + X_{ij}(\lambda^{(1,2)})}$ & 1 \\
 	$(c,\bar{c})$ & $-2(\ell+1)^2$ & $\sqrt{R^{-2}\ell(\ell+2) + X_{ij}(\lambda^{(1,2)})}$ & 0 \\
 	$(A_0,\phi_1,\phi_1^\dagger)_1$ & $(\ell+1)^2$ & $\sqrt{R^{-2}\ell(\ell+2) + X_{ij}(\lambda^{(1,2)})}$ & 0 \\
 	$(A_0,\phi_1,\phi_1^\dagger)_{2,3}$ & $(\ell+1)^2$ & $\sqrt{R^{-2}(\ell+1 \pm R \mu_1)^2 + X_{ij}(\lambda^{(1,2)})}$ & 0 \\
 	$\phi_2$ & $(\ell+1)^2$ & $\sqrt{R^{-2}(\ell+1)^2 + X_{ij}(\lambda^{(1,2)})} \pm \mu_2$ & 0 \\
 	$\phi_3$ & $(\ell+1)^2$ & $\sqrt{R^{-2}(\ell+1)^2 + X_{ij}(\lambda^{(1,2)})} \pm \mu_3$ & 0 \\
 	$(\lambda,\chi_1)$ & $-\ell(\ell+1)$ & $	\sqrt{R^{-2} \left( \ell + \frac{1}{2} \pm \frac{R \mu_1}{2} \right)^2 + X_{ij}(\lambda^{(1,2)})} \pm \frac{\mu_2 + \mu_3}{2}$ & 1 \\
 	$(\chi_2,\chi_3)$ & $-\ell(\ell+1)$ & $	\sqrt{R^{-2} \left( \ell + \frac{1}{2} \pm \frac{R \mu_1}{2} \right)^2 + X_{ij}(\lambda^{(1,2)})} \pm \frac{\mu_2 - \mu_3}{2}$ & 1
\end{tabular}\end{center}
\caption{\footnotesize The energies $\varepsilon_\ell$ for the Higgs branch and gauge group $U(N)$, together with their degeneracies for the various fields. The expressions are valid for vanishing or critical ($\mu_i = R^{-1}$) chemical potentials. All possible sign combinations are allowed. $X_{ij}(\lambda^{(1,2)}) = 4 |\lambda^{(1)}|^2 \sin^2 \left( \frac{i\pi}{n} \right) + 4 |\lambda^{(2)}|^2 \sin^2 \left( \frac{j\pi}{n} \right)$.}
\end{table}

Since
\EQ{
	X_{ij}(\lambda^{(1,2)}) = 4 |\lambda^{(1)}|^2 \sin^2 \left( \frac{i\pi}{N} \right) + 4 |\lambda^{(2)}|^2 \sin^2 \left( \frac{j\pi}{N} \right)
}
appears precisely where $|\varphi|^2$ would appear for $\beta = 0$ and $\mathcal{N}=4$ SYM, we see immediately from \eqref{eq:zeroTeffpot} that at zero temperature the effective potential on the Higgs branch must be independent of $\lambda^{(1)}$ and $\lambda^{(2)}$ and equal to
\EQ{
	V_0 + V_1 = \frac{1}{\textnormal{Vol}(S^3)} \frac{3N^2}{16 R}.
}
At finite temperature, we have
\SP{
	\label{eq:finitetempeffactHiggs}
	V_0 + V_1 = \frac{1}{\textnormal{Vol}(S^3)} \bigg\{
	\frac{3N^2}{16R} -	2T \sum_{i,j=1}^N
		\sum_{k=1}^\infty \frac{\cos(k \alpha_{ij} / T)}{k} \\
		\sum_{l = 1}^\infty l^2 e^{-\frac{k}{T} \sqrt{R^{-2} l^2 + X_{ij}(\lambda^{(1,2)})}} \\ \bigg( 1 + \frac{1}{2} \left[ \cosh \left( \frac{k \mu_2}{T} \right) + \cosh \left( \frac{k \mu_3}{T} \right) \right] - \\ - (-1)^k \left[ \cosh \left( \frac{k (\mu_2 +\mu_3)}{2T} \right) + \cosh \left( \frac{k (\mu_2 - \mu_3)}{2T} \right) \right] \bigg)
	\bigg\}.
}
Again, because of the attractive potential, we can put $\alpha_{ij}=0$ in the above expression, which shows that the large $N$ theory is in the deconfined phase.

Now, let us move on to the more general case when $n$ does not necessarily equal $N$. Again, we will only give VEVs to two of the complex scalar fields:
\EQ{
	\left\langle \phi_1 \right\rangle = \Lambda^{(1)} \otimes U_{(n)}, \ \left\langle \phi_2 \right\rangle = \Lambda^{(2)} \otimes V_{(n)},
}
with
\EQ{
	\Lambda^{(1,2)} = \textnormal{diag} \left( \lambda^{(1,2)}_1, \lambda^{(1,2)}_2, \ldots, \lambda^{(1,2)}_m, \right).
}
We can expand the fluctuations as
\EQ{
	\phi = \frac{1}{\sqrt{2n}} \sum_{a,b=1}^m \sum_{i,j=1}^n \phi_{i,j}^{a,b} M_{a,b} \otimes J_{i,j},
}
where
\EQ{
	\left( M_{a,b} \right)_{de} = \delta_{ad} \delta_{be}
}
is an $m \times m$ matrix.
Then, when the VEVs act on the fluctuations in commutators such as $[\phi_i,\phi]$, instead of getting expressions involving $\sin \theta$ with $\theta = \frac{j\pi}{n}$ or $\theta = \frac{i\pi}{n}$, we will now get expressions of the form
\EQ{
	\lambda^{(i)}_a e^{i\theta} - \lambda^{(i)}_b e^{-i\theta}.
}
It is the absolute value squared which will appear in the expressions for the energy levels. We have
\SP{
	&\left| \lambda^{(i)}_a e^{i\theta} - \lambda^{(i)}_b e^{-i\theta} \right|^2 = \\
	&= \left( \left| \lambda^{(i)}_a \right| - \left| \lambda^{(i)}_b \right| \right)^2 + 4 \left| \lambda^{(i)}_a \right| \left| \lambda^{(i)}_b \right| \sin^2 \theta',
}
with
\EQ{
	\theta' = \frac{1}{2} \left\{ \arg \lambda^{(i)}_a - \arg \lambda^{(i)}_b \right\} + \theta.
}
In other words, nothing is different from the case $n=N$ considered before (and summarized in Table 2) other than that $X$ now takes the form
\SP{
	X_{abij}(\Lambda^{(1,2)}) = \left( \left| \lambda^{(1)}_a \right| - \left| \lambda^{(1)}_b \right| \right)^2 + \left( \left| \lambda^{(2)}_a \right| - \left| \lambda^{(2)}_b \right| \right)^2 + \\ +
	4 \left| \lambda^{(1)}_a \right| \left| \lambda^{(1)}_b \right| \sin^2 \left( \frac{1}{2} \left\{ \arg \lambda^{(1)}_a - \arg \lambda^{(1)}_b \right\} + \frac{i\pi}{n} \right) + \\ +
	4 \left| \lambda^{(2)}_a \right| \left| \lambda^{(2)}_b \right| \sin^2 \left( \frac{1}{2} \left\{ \arg \lambda^{(2)}_a - \arg \lambda^{(2)}_b \right\} + \frac{j\pi}{n} \right).
}
At zero temperature, the one-loop effective action remains the same as for $n=N$ (i.~e. flat), while at finite temperature all that changes in the expression for the one-loop effective potential, equation \eqref{eq:finitetempeffactHiggs}, is the form of $X$ and that we now also have to sum over $a$ and $b$:
\SP{
	\label{eq:generalHiggs}
	V_0 + V_1 = \frac{1}{\textnormal{Vol}(S^3)} \bigg\{
	\frac{3N^2}{16R} -	2T \sum_{a,b=1}^m \sum_{i,j=1}^n
		\sum_{k=1}^\infty \frac{\cos(k \alpha_{ij} / T)}{k} \\
		\sum_{l=1}^\infty l^2 e^{-\frac{k}{T} \sqrt{R^{-2} l^2 + X_{abij}(\Lambda^{(1,2)})}} \\ \bigg( 1 + \frac{1}{2} \left[ \cosh \left( \frac{k \mu_2}{T} \right) + \cosh \left( \frac{k \mu_3}{T} \right) \right] - \\ - (-1)^k \left[ \cosh \left( \frac{k (\mu_2 +\mu_3)}{2T} \right) + \cosh \left( \frac{k (\mu_2 - \mu_3)}{2T} \right) \right] \bigg)
	\bigg\}.
}
The same remarks regarding the differences between gauge group $U(N)$ and $SU(N)$ remain true for the Higgs branch, with the only difference being that in order to use the same expressions for the one-loop effective potential in the two cases, we now need to take the large $m$ limit.

\section{Metastable Phases}
In this section, we will take one or more of the chemical potentials to be near critical, which we define as
\EQ{
	\mu_i = R^{-1} + \mathcal{O}(\lambda),
}
where $\lambda = g^2 N$ is the 't~Hooft coupling. In particular, this means that corrections to the preceding results appear at higher orders in perturbation theory. We will see that even though at a classical level this choice of chemical potential causes an instability, when we take into account the quantum corrections, there are metastable phases at small finite temperature $RT \ll 1$.

First, consider the Coulomb branch at small finite temperature and close to the origin of the moduli space, so that
\EQ{
	R^2 |\varphi_{ij}|^2, R^2 |{\varphi_\beta}_{ij}|^2 \ll RT \ll 1.
}
For $\mu_2 = \mu_3 = 0$, we can put $l=k=1$ in \eqref{eq:finitetempeffact} after which we get that the one-loop quantum correction to the effective potential is given by
\SP{
	V_1 = \frac{1}{\textnormal{Vol}(S^3)} \bigg\{
	\frac{3N^2}{16R} + \frac{R}{2} \sin^2(\pi \beta) \left| \sum_{i=1}^N \varphi_i \right| ^2 - \\ -
	4T \sum_{i,j=1}^N
		\bigg( e^{-\frac{1}{RT} \sqrt{1 + R^2 |\varphi_{ij}|^2}} +
	e^{-\frac{1}{RT} \sqrt{1 + R^2 |{\varphi_\beta}_{ij}|^2}} \bigg)
	\bigg\}.
}
Expanding in $\varphi$ and $\varphi_\beta$, we obtain
\SP{
	V_1 = \frac{1}{\textnormal{Vol}(S^3)} \bigg\{
	\frac{3N^2}{16R} + \frac{R}{2} \sin^2(\pi \beta) \left| \sum_{i=1}^N \varphi_i \right| ^2 - 8T N^2 e^{-\frac{1}{RT}} + \\
	+ 8RN e^{-\frac{1}{RT}} \sum_i |\varphi_i|^2 - 8R e^{-\frac{1}{RT}} \cos^2(\pi \beta) \left| \sum_{i=1}^N \varphi_i \right| ^2 \bigg\},
}
which again is the same result as for the $\mathcal{N}=4$ case in the case of gauge group $SU(N)$, but different for gauge group $U(N)$ \cite{Hollowood:2008gp}.
The tree level term is equal to
\EQ{
	V_0 = \frac{N}{\lambda} (R^{-2} - \mu_1^2) \sum_i |\varphi_i|^2,
}
so we see that we have a metastable state at the origin if
\EQ{
	0 < \mu_1 - R^{-1} < \frac{2 \lambda}{\pi^2 R} e^{-\frac{1}{RT}}.
}
This holds true for gauge group $U(N)$ also, since the only potentially negative contribution to the mass of the new field is suppressed exponentially for $RT \ll 1$. In the large $N$ limit, the decay rate, through tunnelling and thermal activation, becomes zero \cite{Hollowood:2008gp}.

Moving on to the Higgs branch and the case $n=N$, let us put $\mu_3 = 0$, $\alpha_{ij} = 0$. Again, we consider small temperature and VEVs:
\EQ{
	R^2 |\lambda^{(i)}|^2 \ll RT \ll 1.
}
The sum over $k$ in \eqref{eq:finitetempeffactHiggs} contains a piece equal to
\EQ{
	\label{eq:finitetempHiggssum}
	\sum_{k=1}^\infty \frac{1}{k} e^{-\frac{k}{RT} \left( \sqrt{l^2 + 4 R^2 |\lambda^{(1)}|^2 \sin^2 \left( \frac{i\pi}{N} \right) + 4 R^2 |\lambda^{(2)}|^2 \sin^2 \left( \frac{j\pi}{N} \right)} - 1 \right)},
}
which for $l = 1$ clearly leads to a logarithmic divergence for small VEVs. When more than one critical chemical potential is turned on, extra zero modes appear. The sum \eqref{eq:finitetempHiggssum} corresponds precisely to integrating out these zero modes, which really should have been kept in the effective action, and this is what causes the logarithmic divergence near the origin of the moduli space. This is analogous to what happens for $\Ncal = 4$ with two or three critical chemical potentials \cite{Hollowood:2008gp}. The next to leading contribution to the one-loop effective action \eqref{eq:finitetempeffactHiggs} comes from a term which is similar to \eqref{eq:finitetempHiggssum}, with $k = l = 1$ and a $\frac{1}{2}$ instead of a $1$ outside the square root in the exponent. Expanding in $|\lambda^{(i)}|$, we obtain
\SP{
	V_0 + V_1 = \frac{1}{\textnormal{Vol}(S^3)} \bigg\{ \frac{3N^2}{16R} - 4T N^2 e^{-\frac{1}{2RT}} + \\ +
	8 N R \left( \sum_{i=1}^N \sin^2 \left( \frac{i\pi}{N} \right) \right) e^{-\frac{1}{2RT}} \left( |\lambda^{(1)}|^2 + |\lambda^{(2)}|^2 \right) \bigg\}.
}
The only gauge invariant operator consistent with the symmetries of the theory, which would reproduce the same mass squared as above, is proportional to $\Tr ( \phi_1^\dagger \phi_1 + \phi_2^\dagger \phi_2 )$. Therefore, the extra zero modes in the effective action must also have a positive mass squared at the origin of moduli space. This shows that for near (and above) critical chemical potentials there is a metastable state at the origin. In terms of the radii \eqref{eq:torusradii} of the extra-dimensional torus, the torus is metastable at infinite volume, and decays to zero size.

For general $N = n m$, similar considerations lead to a one-loop effective action near the origin of the form
\SP{
	V_0 + V_1 = \frac{1}{\textnormal{Vol}(S^3)} \bigg\{ \frac{3N^2}{16R} - 4T N^2 e^{-\frac{1}{2RT}} + \\ +
	2 N R \sum_{a,b=1}^m \sum_{i,j=1}^n X_{abij}(\Lambda^{(1,2)}) \bigg\},
}
which also has a minimum with positive curvature for zero VEVs, showing that for near (and above) critical chemical potentials there is a metastable state at the origin of moduli space.

\section{Gravity Dual}
\subsection{$AdS_5$ Black Hole Spinning in $S^5$}
Let us first review the Type IIB supergravity solution dual to finite temperature $\mathcal{N}=4$ SYM with chemical potentials. The solution describes an $AdS_5$ black hole spinning in $S^5$. The ten-dimensional background metric is given by \cite{Cvetic:1999xp}
\EQ{
	\label{eq:metric}
	ds^2_{10} = \tilde{\Delta}^{1/2} ds_5^2 + R^2\tilde{\Delta}^{-1/2} \sum_{i=1}^3 X_i^{-1} \left\{ dr_i^2 + r_i^2 \left( d\phi_i + R^{-1} A_i^{(1)} \right)^2 \right\},
}
where
\EQ{
	ds_5^2 = - H(r)^{-2/3} f(r) dt^2 + H(r)^{1/3} [ f(r)^{-1} dr^2 + r^2 d\Omega_{3,1}^2 ]
}
is the metric of the $AdS_5$ black hole, and $d\Omega_{3,1}$ is the volume element of the $S^3$. We have
\AL{
	H_i(r) &= 1 + \frac{q_i}{r^2}, \\
	H(r) &= H_1(r) H_2(r) H_3(r), \\
	f(r) &= 1 - \left( \frac{r_0}{r} \right)^2 + \left( \frac{r}{R} \right)^2 H(r), \\
	r_0 &= r_H \left( 1 + \left( \frac{r_H}{R} \right)^2 H(r_H) \right)^{1/2}, \\
	X_i &= H(r)^{1/3}/H_i(r), \\
	A_{i\mu} &= -\frac{e_i}{r^2 + q_i} \delta_{\mu,0}, \\
	e_i &= \sqrt{q_i(r_0^2 + q_i)}, \\
	\tilde \Delta &= \sum_{i=1}^3 X_i r_i^2, \\
	\sum_i r_i^2 &= 1.
}
In addition to the metric, we have the self-dual five-form $F^{(5)} = dc_4 = d\tilde c_4 + * d\tilde c_4$ with \cite{Yamada:2008em}
\SP{
	\tilde c_4 = \left[ \left( \frac{r}{R} \right)^4 \Delta -  \sum_i \frac{r_0^2 + (-r_H^2 + q_i)}{R^2} r_i^2 \right] dt \wedge \epsilon^{(3)} + \\ + \sum_i \left( \frac{e_i}{R^2} \right) r_i^2 (R d\phi_i)\wedge \epsilon ^{(3)},
}
where $\Delta \equiv H^{2/3} \tilde{\Delta}$, and $\epsilon ^{(3)}$ is the volume form with respect to $R^2 d\Omega_{3,1}$.

After going to the co-rotating frame
\EQ{
	\phi_i \rightarrow \phi_i - R^{-1} A_{i0}(r_H) t,
}
in which the horizon of the black hole is static, the only change in the metric \eqref{eq:metric} is
\EQ{
	A_{i0} \rightarrow \frac{e_i}{r_H^2 + q_i} - \frac{e_i}{r^2 + q_i}.
}
Also, the new expression for $d\tilde c_4$ is
\SP{
	\tilde c_4 = \left[ \left( \frac{r}{R} \right)^4 \Delta + \sum_i \frac{1}{R^2} \left\{ \frac{e_i^2}{r_H^2 + q_i} - (r_0^2 - r_H^2 + q_i) \right\} r_i^2 \right] dt \wedge \epsilon^{(3)} + \\ + \sum_i \left( \frac{e_i}{R^2} \right) r_i^2 (R d\phi_i)\wedge \epsilon ^{(3)}.
}
We can identify the chemical potentials of the field theory on the boundary as
\EQ{
	\mu_i = A_{i0}(\infty)/R = R^{-1} \frac{e_i}{r_H^2 + q_i}.
}

\subsection{The $\beta$-deformed Solution}
The idea of Lunin and Maldacena \cite{Lunin:2005jy} was to obtain the Type IIB background describing the $\beta$-deformed theory by performing a TsT-trans-formation (see Appendix) on the solution describing $\mathcal{N}=4$ SYM. Starting with the solution in $AdS_5 \times S^5$
\AL{
	\label{eq:beforeLM}
	ds^2 &= ds^2_{AdS_5} + R^2 \sum_{i=1}^3 (dr_i^2 + r_i^2d\phi_i^2), \\
	C_4 &= \omega_4 + 4 R^4 \omega_1 \wedge d\phi_1 \wedge \phi_2 \wedge \phi_3, \\
	e^{2\phi} &= e^{2\phi_0},
}
then going to coordinates
\AL{
	\phi_1 &= \varphi_3 - \varphi_2, \\
	\phi_2 &= \varphi_1 + \varphi_2 + \varphi_3, \\
	\phi_3 &= \varphi_3 - \varphi_1,
}
and performing a T-duality along $\varphi_1$, followed by a small shift $\varphi_2 \rightarrow \varphi_2 + \gamma \varphi_1$, ($\gamma \equiv \beta$) and then another T-duality along $\varphi_1$,
they obtained the TsT-transformed Lunin-Maldacena solution
\ALlabel{
	ds^2 &= ds^2_{AdS_5} + R^2 \left[ \sum_{i=1}^3 (dr_i^2 + G r_i^2d\phi_i^2) + \hat \gamma^2 G r_1^2 r_2^2 r_3^2 \left( \sum_{i=1}^3 d\phi_i \right)^2 \right], \\
	B^{NS} &= R^2 \hat \gamma G (r_1^2 r_2^2 d\phi_1 \wedge d\phi_2 + r_2^2 r_3^2 d\phi_2 \wedge d\phi_3 + r_3^2 r_1^2 d\phi_3 \wedge d\phi_1), \\
	C_2 &= -4 R^2 \hat \gamma \omega_1 \wedge (d\phi_1 +d\phi_2 + d\phi_3), \\
	C_4 &= \omega_4 + 4 G R^4 \omega_1 \wedge d\phi_1 \wedge \phi_2 \wedge \phi_3, \\
	e^{2\phi} &= G e^{2\phi_0},
}{eq:LM}
where
\AL{
	G^{-1} &= 1 + \hat \gamma (r_1^2 r_2^2 + r_2^2 r_3^2 +r_3^2 r_1^2), \\
	\hat \gamma &= R^2 \gamma \equiv R^2 \beta,
}
and
\AL{
	r_1 &= \cos \alpha, \\
	r_2 &= \sin \alpha \cos \theta, \\
	r_3 &= \sin \alpha \sin \theta, \\
	d\omega_1 &= \cos \alpha \sin^3 \alpha \sin \theta \cos \theta d\alpha \wedge d\theta, \\
	d\omega_4 &= \omega_{AdS_5}.
}

In order to obtain the correct background for the $\beta$-deformed theory at finite temperature and with chemical potentials, we should perform a TsT-transformation on the solution given in the previous section. First, we note that a coordinate change $\phi_i \rightarrow \phi_i + v t$ followed by TsT is the same as vice versa. (This follows directly from the form of the transformation rules given in the Appendix: as long as a coordinate transformation does not mix the two coordinates along which we T-dualize with each other, the TsT-transformed expressions behave as tensors.) It is convenient to make the coordinate change
\EQ{
	\phi_i' = \phi_i + R^{-1} A_{i0} t,
}
after which, apart from a few scaling factors, the metric is the same as in \eqref{eq:beforeLM}. Only the components of the metric and $B^{NS}$ involving the coordinates $\phi_i$ are affected by the TsT-transformation. This means that we can take the LM solution \eqref{eq:LM} for the metric and $B^{NS}$ and simply make the following substitutions
\AL{
	R &\rightarrow \tilde{\Delta}^{-1/4} R, \\
	A^{(1)} &\rightarrow \tilde{\Delta}^{-1/4} A^{(1)}, \\
	r_i &\rightarrow X_i^{-1/2} r_i,
}
to obtain the correct form of the Type IIB supergravity solution describing $\beta$-deformed $\mathcal{N}=4$ SYM at finite temperature with chemical potentials:
\AL{
	ds^2_{10} &= \tilde{\Delta}^{1/2} ds_5^2 + R^2\tilde{\Delta}^{-1/2} \Bigg[ \sum_{i=1}^3 X_i^{-1} \left\{ dr_i^2 + G r_i^2 d\phi_i'^2 \right\} + \nonumber \\ & +
	\hat{\gamma}^2 G \frac{r_1^2 r_2^2 r_3^2}{X_1 X_2 X_3} \left( \sum_{i=1}^3 d\phi_i' \right)^2 \Bigg], \\
	B^{NS} &= \hat{\gamma} G R^2 \tilde{\Delta}^{-1/2} \bigg( \frac{r_1^2 r_2^2}{X_1 X_2} d\phi_1' \wedge d\phi_2' + \nonumber \\
	&+ \frac{r_2^2 r_3^2}{X_2 X_3} d\phi_2' \wedge d\phi_3' + \frac{r_3^2 r_1^2}{X_3 X_1} d\phi_3' \wedge d\phi_1' \bigg), \\
	e^{2\phi} &= G e^{2\phi_0},
}
where
\AL{
	G^{-1} &= 1+\hat{\gamma}^2 \left( \frac{r_1^2 r_2^2}{X_1 X_2} + \frac{r_2^2 r_3^2}{X_2 X_3} + \frac{r_3^2 r_1^2}{X_3 X_1}  \right), \\
	\hat{\gamma} &\equiv R^2 \tilde{\Delta}^{-1/2} \gamma.
}
The TsT-transformed $n$-forms $C_n$ can be found from the formula \cite{Imeroni:2008cr}
\EQ{
    \label{eq:TsTnform}
    \sum_q C_q \wedge e^{-B} = \sum_q c_q \wedge e^{-b} + \gamma \left[ \sum_q c_q \wedge e^{-b}
\right]_{[\varphi^1][\varphi^2]},
}
where $c_q$ are the untransformed $n$-forms, b is the untransformed NS 2-form, and for a general $p$-form $\omega_p$ we have defined
\EQ{
	\omega_p = \bar \omega_p + \omega_{p[y]} \wedge dy,
}
where $\bar \omega_p$ does not contain any legs in $dy$. \eqref{eq:TsTnform} is to be understood as being valid order by order. In particular, for our solution we have that
\ALlabel{
	C_0 &= 0, \\
	C_2 &= \gamma \left[ c_4 \right]_{[\varphi^1][\varphi^2]}, \\
	C_4 - C_2 \wedge B &= c_4, \\
	C_6 - C_4 \wedge B &= 0, \\
	C_8 &= 0,
}{eq:TsTnformexpl}
where we have used that $b=0$ and $B \wedge B = 0$.

\section{Probe-Brane Calculation}
\subsection{Coulomb Branch}
We will now perform a probe-brane calculation in the TsT-transformed background. The Coulomb branch of the theory is probed by a D3-brane, static in the co-rotating frame (in which the horizon of the black hole also is static), and extending in all the directions of $AdS_5$ except the radial direction \cite{Kraus:1998hv}. In the field theory, separating a D3-brane from the stack of $N$ branes at the origin, corresponds to turning on VEVs
\EQ{
	\phi_i = \textnormal{diag} (v_i,-\frac{v_i}{N-1}, \cdots, -\frac{v_i}{N-1}).
}
The action for a general D$p$-brane has the form
\SP{
	S_{\text{D}p} = - \tau_p \int d^{p+1}\sigma \ e^{-\phi}
		\sqrt{- \det \left(\hat{G}_{ab} + F_{ab} - \hat{B}_{ab}\right) }
		 - \\ - \tau_p \int_{\mathcal{M}_{p+1}} \sum_q \hat{C}_q \wedge e^{F-\hat{B}},
}
where $\tau_p = \frac{1}{(2\pi)^p g_s}$, and hats denote pullbacks onto the world-volume $\mathcal{M}_p$ of the brane. For the D3-brane, $F = \hat B = 0$, and the induced metric is given by
\SP{
	\tilde{\Delta}^{1/2} \left( - H(r)^{-2/3} f(r) dt^2 + H(r)^{1/3} r^2 d\Omega_{3,1}^2 \right)
	+ \\ + \tilde{\Delta}^{-1/2} \Bigg\{ \sum_{i=1}^3 X_i^{-1} G r_i^2  A_{i0}^2 + \hat{\gamma}^2 G \frac{r_1^2 r_2^2 r_3^2}{X_1 X_2 X_3} \left( \sum_{i=1}^3 A_{i0} \right)^2 \Bigg\} dt^2,
}
so that
\SP{
	e^{-\phi} \sqrt{- \det \left(\hat{G}_{ab}\right) } = e^{-\phi_0} r^3 \tilde{\Delta} \Bigg\{ G^{-1} H^{1/3} f - \\ - \tilde{\Delta}^{-1} H \Bigg[ \sum_{i=1}^3 X_i^{-1} r_i^2  A_{i0}^2 + \hat{\gamma}^2 \frac{r_1^2 r_2^2 r_3^2}{X_1 X_2 X_3} \left( \sum_{i=1}^3 A_{i0} \right)^2 \Bigg] \Bigg\}^{1/2}.
}
For the Wess-Zumino term, we have
\SP{
	\label{eq:WZD3}
	\int_{\mathcal{M}_4} ( \hat{C}_4 - \hat{C}_2 \wedge \hat{B} ) = \int_{\mathcal{M}_4} \hat{c}_4 = \\
	= \int_{\mathcal{M}_4} \left[ \left( \frac{r}{R} \right)^4 \Delta + \sum_i \frac{1}{R^2} \left\{ \frac{e_i^2}{r_H^2 + q_i} - (r_0^2 - r_H^2 + q_i) \right\} r_i^2 \right] dt \wedge \epsilon^{(3)},
}
which is the same as for the $\Ncal = 4$ case analyzed in \cite{Yamada:2008em}.

First we note that, at the horizon, the terms that are introduced by the deformation have no dependence on any of the coordinates parameterizing the $S^5$; this is because $f(r_H) = A_{i0}(r_H) = 0$. If we turn on just one chemical potential $\mu_2 = \mu_3 = 0$, then the probe-brane action is minimized for $r_1=1$, $r_2 = r_3 = 0$, in which case $G = 1$, and all $\gamma$-dependence disappears. Therefore, the analysis is exactly the same as for the undeformed case; for close to but above critical chemical potential, there will be a metastable state at $r = r_H$, which decays towards the run-away direction $r = \infty$ \cite{Yamada:2008em}. For two equal chemical potentials ($\mu_1 = \mu_2$, $\mu_3 = 0$), the undeformed probe-brane action is minimized for $r_3 = 0$, but has no dependence on $r_1$ or $r_2$. Since no such dependence is introduced at the horizon by the $\beta$-deformation, there is still a meta-stable state at $r = r_H$. For three equal critical chemical potentials, the probe-brane action has no dependence on either of the coordinates $r_i$ in the undeformed case. Again, at the horizon, no such dependence is introduced by the $\beta$-deformation. We note that a probe-brane at the black hole horizon $r = r_H$ should correspond to zero VEVs in the field theory.

\subsection{Higgs Branch}
\begin{figure}[t]
	\centering
		\includegraphics[width=13cm]{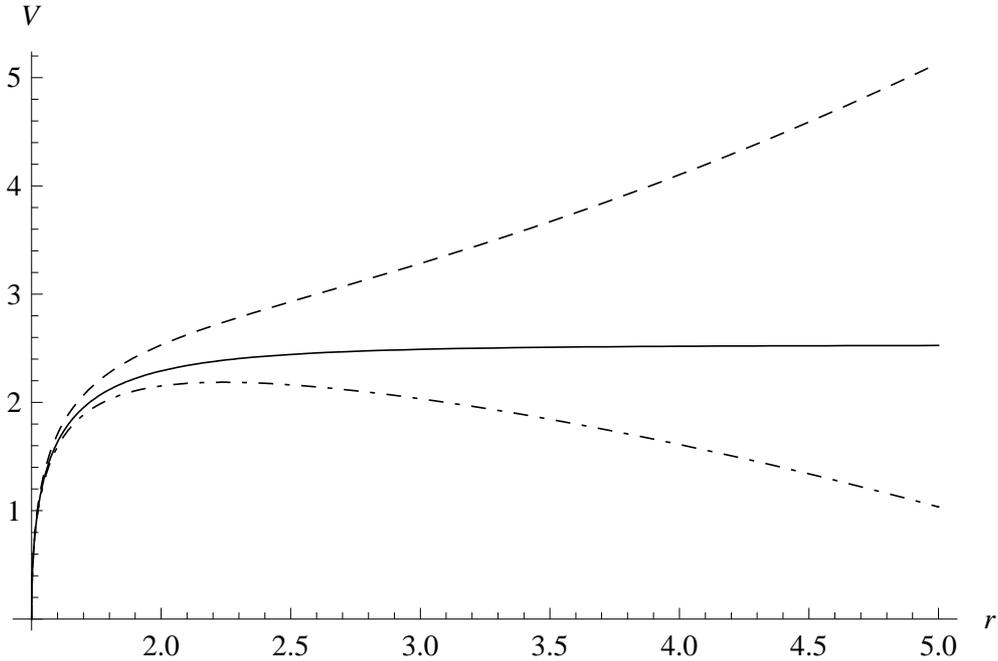}
	\caption{\footnotesize The D5 probe-brane action $V$ for chemical potentials $\mu_i = (q,q,q)$ as a function of the radius $r$. Up to a factor of $\frac{1}{n}$, this is the same as the action of a D3 probe-brane (which does not wrap the torus) in the undeformed background. Everything is in units of $R$. The solid line corresponds to critical chemical potential $q = 1$, the dashed line corresponds to $q = 0.7$, and the dot-dashed line corresponds to $q = 1.2$. In all cases, we have put $r_H = 1.5$.}
	\label{fig:plot}
\end{figure}
The Higgs branch is probed by a D5-brane extending in the same directions as the D3-brane of the Coulomb branch, but in addition wrapping the torus formed by the two coordinates of the $S^5$ that are involved in the TsT-transformation \cite{Dorey:2003pp, Dorey:2004iq}. In the field theory, this corresponds to VEVs given by
\AL{
	\left\langle \phi_1 \right\rangle &= \Lambda^{(1)} \otimes U_{(n)}, \\
	\left\langle \phi_2 \right\rangle &= \Lambda^{(2)} \otimes V_{(n)}, \\
	\left\langle \phi_3 \right\rangle &= \Lambda^{(3)} \otimes V_{(n)}^\dagger U_{(n)}^\dagger,
}
with
\EQ{
	\Lambda^{(i)} = \textnormal{diag} (v^{(i)},-\frac{v^{(i)}}{m-1}, \cdots, -\frac{v^{(i)}}{m-1}).
}
Also, we will need to turn on a world-volume flux along the directions of the torus:
\EQ{
    F_{\varphi_1 \varphi_2} = \frac{1}{\gamma}.
}
One way of seeing this is that the D3-brane RR-charge of one D5-brane should be the same as that of $n = 1/\gamma$ D3-branes. Using \eqref{eq:TsTnformexpl} and $F \wedge F = B \wedge B = 0$, the Wess-Zumino term is
\SP{
	\int_{\mathcal{M}_6} \left( \hat{C}_6 + \hat{C}_4 \wedge (F - \hat{B}) + \frac{1}{2} \hat{C}_2 \wedge (F - \hat{B})^2 \right) = \int_{\mathcal{M}_6} \hat{c}_4 \wedge F,
}
which indeed is equal to $n$ times the corresponding expression \eqref{eq:WZD3} for a D3-brane, which in turn is the same as that for a D3-brane in the undeformed background corresponding to $\Ncal = 4$ studied in \cite{Yamada:2008em}.

For the world-volume part of the action, a more involved calculation gives
\SP{
	e^{-\phi} \sqrt{- \det \left(\hat{G}_{ab} + F_{ab} - \hat{B}_{ab}\right) } = \\ = e^{-\phi_0} \gamma^{-1} r^3 \tilde{\Delta} \left( H^{1/3} f - \tilde{\Delta}^{-1} H \sum_{i=1}^3 X_i^{-1} r_i^2  A_{i0}^2 \right)^{1/2},
}
which also is precisely equal to $n$ times the corresponding result for a D3-brane in the undeformed background. Therefore all the results of \cite{Yamada:2008em} apply in the TsT-transformed case. In particular, for nearly critical chemical potentials, there is a metastable state with a D5-brane situated at $r = r_H$, which will eventually be ``ejected'' towards infinite radius.

\section{Conclusions}
We have studied the $\beta$-deformation of $\mathcal{N}=4$ SYM on $S^3$ with chemical potentials. On the Coulomb branch, the one-loop effective potential at zero temperature and critical chemical potentials is flat for gauge group $SU(N)$, but for $U(N)$, there is a dependence on the overall $U(1)$ traceful part of the VEV. On the Higgs branch, the zero temperature one-loop effective action is flat both for $SU(N)$ and $U(N)$. This is expected since on the Higgs branch, the low energy theory is $\Ncal = 4$, and can be viewed as a six-dimensional theory with 16 supercharges compactified on a torus. At near critical chemical potential and small finite temperature, there is a metastable state at the origin of moduli space for both the Coulomb branch and Higgs branch. On the Higgs branch, this has the interpretation as an extra-dimensional torus which becomes metastable for infinite size and decays to zero size through quantum tunnelling and thermal activation. We have found the Type IIB supergravity background which describes the theory at strong 't~Hooft coupling. At finite temperature, this solution describes a black hole rotating in the internal $S^5$. The Coulomb branch is probed by a D3-brane, whereas the Higgs branch is probed by a D5-brane wrapping a torus. On both the Coulomb branch and the Higgs branch, for near (and above) critical chemical potentials there are metastable states in which the probe-branes reside at the black hole horizon and tunnel out towards infinite radius. This matches the weak coupling picture.

\section*{Acknowledgements}
We would like to thank E.~Imeroni, T.~Hollowood, and especially S.~P.~Kumar and A.~Naqvi for useful discussions. This work was in part supported by a grant from Stiftelsen Karl och Annie Leons Minnesfond för Vetenskaplig forskning.

\appendix

\section{Conventions and Gauge Fixing}
We normalize the generators of the $SU(N)$ ($U(N)$) Lie algebra as follows:
\EQ{
	\Tr \, T^a T^b = \frac{1}{2} \delta_{ab}.
}
This implies (for U(N) the second term on the right hand side is not present)
\EQ{
	T^a_{ij} T^a_{kl} = \frac{1}{2} \delta_{il} \delta_{jk} - \frac{1}{2N} \delta_{ij} \delta_{kl},
}
which in turn implies that (again, the second term is not present for $U(N)$)
\EQ{
	\label{eq:superderivation}
	\left( \Tr \, X T^a \right) \left( \Tr \, T^a Y \right) = \frac{1}{2} \Tr \, XY - \frac{1}{2N} \left( \Tr \, X \right) \left( \Tr \, Y \right).
}
The superpotential of the $\beta$-deformed theory is
\EQ{
	W = i 2 \sqrt{2} \, \Tr \, \Phi_1 [\Phi_2,\Phi_3]_\beta.
}
In the following, we will focus on the $U(N)$ case. Using \eqref{eq:superderivation}, the potential for the scalars coming from the superpotential is
\EQ{
	V_W = \frac{4}{g^2} \, \Tr \left( | [\phi_1,\phi_2]_\beta |^2 + | [\phi_2,\phi_3]_\beta |^2 + | [\phi_3,\phi_1]_\beta |^2 \right).
}
The potential due to the D-term is
\EQ{
	V_D = \frac{1}{g^2} \Tr \left( [\phi_1^\dagger,\phi_1] + [\phi_2^\dagger,\phi_2] + [\phi_3^\dagger,\phi_3] \right)^2
}
Adding chemical potentials to the theory modifies the kinetic terms of the scalars to
\SP{
	2 \Tr \left( (D_\mu + \mu_i \delta_{0,\mu}) \phi_i \right)^\dagger (D_\mu - \mu_i \delta_{0,\mu}) \phi_i = \\ =
	2 \Tr \left( (D_\mu \phi_i)^\dagger D_\mu \phi + 2 \mu_i \phi_i^\dagger \tilde{D}_0 \phi_i - \mu_i^2 \phi_i^\dagger \phi_i \right)
}
There are also conformal mass terms
\EQ{
	2 \Tr \, R^{-2} \phi_i^\dagger \phi_i,
}
which for critical values $\mu_i = R^{-1}$ of the chemical potentials cancel against the negative mass squared terms generated by the chemical potentials, thus opening up flat directions. We give VEVs for potentially flat directions
\EQ{
	\phi_i \rightarrow \varphi_i + \phi_i
}
These satisfy the F-term and D-term equations:
\AL{
	[\varphi_1,\varphi_2]_\beta = [\varphi_2,\varphi_3]_\beta = [\varphi_3,\varphi_1]_\beta = 0, \\
	\sum_{i=1}^3 [\varphi_i^\dagger,\varphi_i] = 0.
}
Using the Jacobi identity and the D-term equation, we have that
\EQ{
	\sum_{i=1}^3 \varphi_i^\dagger \varphi_i = \sum_{i=1}^3 \varphi_i \varphi_i^\dagger
}
In order to fix the gauge, we add a term
\EQ{
	\mathcal{L}_{gf} = \frac{1}{g^2} \Tr \left( \nabla_i A^i + \tilde{D}_0 A^0 - i \sum_{i=1}^3 \left( [\varphi_i^\dagger,\phi_i] + [\varphi_i,\phi_i^\dagger] \right) \right)^2
}
to the Lagrangian, corresponding to $R_\xi$-gauge with Feynman parameter $\xi=1$. This cancels cross terms of the form
\EQ{
	\frac{1}{g^2} \left( - i [A_\mu,\varphi_i^\dagger] \partial_\mu \phi_i
	- i \partial_\mu \phi_i^\dagger [A_\mu,\varphi_i] \right).
}
The kinetic terms for the fields with a critical chemical potential have the following form:
\EQ{
	2 \Tr \left( (D_\mu (\varphi_i + \phi_i))^\dagger D_\mu (\varphi_i + \phi_i) + 2 \mu_i (\varphi_i + \phi_i)^\dagger \tilde{D}_0 (\varphi_i + \phi_i) \right)
}
Taking care to cancel the cross terms from the gauge fixing, the first term contributes
\EQ{
	2 \Tr \left( \phi_i^\dagger (-D^2) \phi_i + \frac{1}{2} A_\mu (2 \varphi_i^\dagger \varphi_i) A_\mu \right)
}
at second order, whereas the second term contributes
\EQ{
	2 \Tr \left( 2 \mu_i \left[ \phi_i^\dagger \tilde{D}_0 \phi_i
	+ i \phi_i^\dagger \varphi_i A_0 + i A_0 \varphi_i^\dagger \phi_i \right] \right),
}
where we have used the following notation for the commutator action:
\AL{
	\varphi &\equiv [\varphi, \cdot] \\
 	\varphi_\beta &\equiv [\varphi, \cdot]_\beta, \\
 	\varphi_\beta^\dagger &\equiv [\varphi^\dagger, \cdot]_{-\beta}.
}
Some useful relations are:
\AL{
	\Tr \, [X,A] B &= - \Tr \, A [X,B], \\
	[A,B]_\beta &= - [B,A]_{-\beta}, \\
	[A,B]_\beta^\dagger &= - [A^\dagger,B^\dagger]_\beta, \\
	\Tr \, [X,A]_\beta^\dagger B &= - \Tr \, A^\dagger [X^\dagger,B]_{-\beta}.
}

\section{Energies For One VEV}

\subsection{Scalars}

Put $\mu_1 = R^{-1}$ and $\mu_2 = \mu_3 = 0$, and give a diagonal VEV to one scalar:
\EQ{
	\varphi_1 = \varphi, \ \varphi_2 = \varphi_3 = 0.
}
At second order, the contribution from the D-term comes from
\EQ{
	V_D^{(2)} = \frac{1}{g^2} \Tr \left( [\varphi^\dagger,\phi_1] - [\varphi,\phi_1^\dagger] \right)^2,
}
whereas the gauge fixing contributes
\EQ{
	V_{gf}^{(2)} = - \frac{1}{g^2} \Tr \left( [\varphi^\dagger,\phi_1] + [\varphi,\phi_1^\dagger] \right)^2.
}
Together, this becomes
\EQ{
	V_D^{(2)} + V_{gf}^{(2)} = - \frac{4}{g^2} \Tr \, [\varphi,\phi_1^\dagger] [\varphi^\dagger,\phi_1],
}
which we write using the commutator action notation as
\EQ{
	V_D^{(2)} + V_{gf}^{(2)} = \frac{1}{g^2} 2 \Tr \, \phi_1^\dagger ( 2 \varphi \varphi^\dagger ) \phi_1.
}
At second order, the superpotential contributes
\EQ{
	V_W^{(2)} = \frac{4}{g^2} \, \Tr \left( | [\varphi,\phi_2]_\beta |^2 + | [\phi_3,\varphi]_\beta |^2 \right),
}
which in commutator action notation is
\EQ{
	V_W^{(2)} = \frac{1}{g^2} 2 \Tr \left( \phi_2^\dagger ( 2 \varphi_\beta^\dagger \varphi_\beta ) \phi_2 + \phi_3^\dagger ( 2 \varphi_{-\beta}^\dagger \varphi_{-\beta} ) \phi_3 \right).
}
Finally, after rescaling $\varphi \rightarrow \varphi/\sqrt{2}$, the bosonic part of the Lagrangian at second order is equal to
\SP{
	V_b^{(2)} = \frac{1}{g^2} 2 \Tr \bigg( \frac{1}{2} A_0 ( - \tilde{D}_0^2 - \nabla^{(s)} + \varphi^\dagger \varphi ) A_0 + \\
	+ \frac{1}{2} A_i ( - \tilde{D}_0^2 - \nabla^{(v)} + \varphi^\dagger \varphi ) A_i + \\
	+ \bar{c} ( - \tilde{D}_0^2 - \nabla^{(s)} + \varphi^\dagger \varphi ) c + \\
	+ \phi_1^\dagger (-\tilde{D}_0^2 - \nabla^{(s)} + \varphi^\dagger \varphi ) \phi_1 + \\
	+ \mu_1 \left[ 2 \phi_1^\dagger \tilde{D}_0 \phi_1 + i \sqrt{2} \phi_1^\dagger \varphi A_0 + i \sqrt{2} A_0 \varphi^\dagger \phi_1 \right] + \\
	+ \phi_2^\dagger ( - \tilde{D}_0^2 - \nabla^{(s)} + \varphi_\beta^\dagger \varphi_\beta + R^{-2} ) \phi_2 + \\
	+ \phi_3^\dagger ( - \tilde{D}_0^2 - \nabla^{(s)} + \varphi_{-\beta}^\dagger \varphi_{-\beta} + R^{-2} ) \phi_3
	\bigg),
}
where we have also included the ghost fields $(\bar c, c)$. The fluctuation matrix for $(A_0,\phi_1,\phi_1^\dagger)$ is:

\EQ{
	{\scriptsize
	\begin{pmatrix}
		- \tilde{D}_0^2 - \nabla^{(s)} + \varphi^\dagger \varphi & i \sqrt{2} R^{-1} \varphi^\dagger & - i \sqrt{2} R^{-1} \varphi \\
		i \sqrt{2} R^{-1} \varphi & - \tilde{D}_0^2 - \nabla^{(s)} + \varphi^\dagger \varphi + 2 R^{-1} \tilde{D}_0 & 0 \\
		- i \sqrt{2} R^{-1} \varphi^\dagger & 0 & - \tilde{D}_0^2 - \nabla^{(s)} + \varphi^\dagger \varphi - 2 R^{-1} \tilde{D}_0 
	\end{pmatrix}}.
}
Putting its determinant
\EQ{
	(- \tilde{D}_0^2 + \ell(\ell+2) R^{-2} + \varphi^\dagger \varphi)
	(- \tilde{D}_0^2 + \ell^2 R^{-2} + \varphi^\dagger \varphi)
	(- \tilde{D}_0^2 + (\ell+2)^2 R^{-2} + \varphi^\dagger \varphi)
}
equal to zero and solving for $\tilde{D}_0$ gives the three energy levels associated with $(A_0,\phi_1,\phi_1^\dagger)$. Here we have expanded the fields in spherical harmonics and used that
\EQ{
	\Delta^{(s)} Y_{\ell} = R^{-2} \ell(\ell+2) Y_\ell \ \ \ \ (\ell=0,1,\ldots).
}

\subsection{Fermions}
The fermions $\chi_i$ couple to the gaugino as
\EQ{
	\frac{1}{g^2} 2 \Tr \left( - i \sqrt{2} \lambda [\phi_i^\dagger,\chi_i] \right) + c.c.
}
There is also a contribution from superpotential, equal to
\EQ{
	- \frac{1}{g^2} 2 i \sqrt{2} \Tr \left( \chi_1 [\phi_2,\chi_3]_\beta + \chi_2 [\phi_3,\chi_1]_\beta + \chi_3 [\phi_1,\chi_2]_\beta  \right) + c.c.
}
The second order Lagrangian for the fermions is
\SP{
	\mathcal{L}_f = 2 \Tr \bigg( \sum_{i=0}^3 \bar{\chi_i} ( i \sigma_\mu D^\mu - i \bar{\mu_i} ) \chi_i -
	\chi_0 (i \varphi^\dagger) \chi_1 - \bar{\chi}_1 ( -i \varphi) \bar{\chi}_0 - \\
	- \chi_3 (i \varphi_\beta) \chi_2 - \bar{\chi}_2 ( -i \varphi_\beta^{\dagger}) \bar{\chi}_3 \bigg),
}
where $\chi_0 \equiv \lambda$, and
\SP{
	\bar\mu_0 &= \tfrac{1}{2} (\mu_1 + \mu_2 + \mu_3) \\
	\bar\mu_1 &= \tfrac{1}{2} (\mu_1 - \mu_2 - \mu_3) \\
	\bar\mu_2 &= \tfrac{1}{2} (-\mu_1 + \mu_2 - \mu_3) \\
	\bar\mu_3 &= \tfrac{1}{2} (-\mu_1 - \mu_2 + \mu_3).
}
We will evaluate the determinant of the fluctuation matrix by brute force. In the path integral, we will expand $e^{-S}$ to the power that saturates the measure
\EQ{
	\int \prod_{i=0}^3 D \bar{\chi}_i(p) D \bar{\chi}_i(-p) D \chi_i(p) D \chi_i(-p).
}
This happens for $S$ to the 16th power. However, matters simplify, because of the block diagonal form of the fluctuation matrix, and also because of how the $\chi_i(p)$s and $\chi_i(-p)$s have to combine. We can represent the way the terms in the Lagrangian combine as four graphs that look like:
\SP{
	\begin{matrix}
		\chi_0(-p) & \leftarrow (-i \varphi) \rightarrow & \chi_1(p) \\
		\uparrow & & \uparrow \\
		(i \sigma_\mu D^\mu(-p) - i \bar{\mu_0}) & & (i \sigma_\mu D^\mu(p) - i \bar{\mu_1}) \\
		\downarrow & & \downarrow \\
		\bar{\chi}_0(p) & \leftarrow (i \varphi) \rightarrow & \bar{\chi}_1(-p)
	\end{matrix}
}
There is another graph which is the same, but with $p \rightarrow -p$, and similarly for $\chi_2$ and $\chi_3$ (but with $\varphi \rightarrow \varphi_\beta$).\footnote{When all three VEVs are turned on, there is a single graph which is a four-dimensional hypercube. When two VEVs are turned on, this gets cut into two cubes, which then get cut into the four squares for one VEV.} Since the four graphs do not connect, we can consider them separately. Following closed paths around the graphs, there are three ways to saturate the measure; the two closed paths
\SP{
	\begin{matrix}
		\chi_0(-p) & & \chi_1(p) \\
		\uparrow \downarrow & & \uparrow \downarrow \\
		\bar{\chi}_0(p) &  & \bar{\chi}_1(-p)
	\end{matrix}
}
pick up a term
\SP{
	\frac{1}{4} (i \sigma_\mu D^\mu(-p) - i \bar{\mu_0})^{\dot{\alpha}\alpha} (i \sigma_\mu D^\mu(-p) - i \bar{\mu_0})_{\dot{\alpha}\alpha} \\
	(i \sigma_\nu D^\nu(p) - i \bar{\mu_1})^{\dot{\beta}\beta} (i \sigma_\nu D^\nu(p) - i \bar{\mu_1})_{\dot{\beta}\beta} = \\
	= ( \tilde{D}_0^2 + \nabla^{(f)} + 2 \bar{\mu}_0 \tilde{D}_0 + \bar{\mu}_0^2) ( \tilde{D}_0^2 + \nabla^{(f)} - 2 \bar{\mu}_1 \tilde{D}_0 + \bar{\mu}_1^2),
}
while the two closed paths
\SP{
	\begin{matrix}
		\chi_0(-p) & \rightleftharpoons & \chi_1(p) \\
		& & \\
		\bar{\chi}_0(p) & \rightleftharpoons & \bar{\chi}_1(-p)
	\end{matrix}
}
pick up
\EQ{
	(\varphi^\dagger \varphi)^2,
}
and, finally, the one closed path that travels around the whole graph
\SP{
	\begin{matrix}
		\chi_0(-p) & \rightarrow & \chi_1(p) \\
		\uparrow & & \downarrow \\
		\bar{\chi}_0(p) & \leftarrow & \bar{\chi}_1(-p)
	\end{matrix}
}
picks up
\SP{
	\frac{1}{4} (i \sigma_\mu D^\mu(-p) - i \bar{\mu_0})^{\dot{\alpha}\alpha} (i \sigma_\nu D^\nu(p) - i \bar{\mu_1})^{\dot{\beta}\beta} (i \varphi^\dagger \epsilon_{\dot{\alpha}\dot{\beta}}) (-i \varphi \epsilon_{\alpha\beta}) = \\
	= 2 \varphi^\dagger \varphi ( - \tilde{D}_0^2 - \nabla^{(f)} + (\tilde{\mu_1} - \tilde{\mu_0}) \tilde{D}_0 + \tilde{\mu_0} \tilde{\mu_1}).
}
Together, we have
\SP{
	( \tilde{D}_0^2 + \nabla^{(f)} + 2 \bar{\mu}_0 \tilde{D}_0 + \bar{\mu}_0^2) ( \tilde{D}_0^2 + \nabla^{(f)} - 2 \bar{\mu}_1 \tilde{D}_0 + \bar{\mu}_1^2) - \\
	+ 2 \varphi^\dagger \varphi ( - \tilde{D}_0^2 - \nabla^{(f)} + (\tilde{\mu_1} - \tilde{\mu_0}) \tilde{D}_0 + \tilde{\mu_0} \tilde{\mu_1}) + (\varphi^\dagger \varphi)^2.
}
Putting this equal to zero and solving for $\tilde{D}_0$ yields the following energies ($\nabla^{(f)} = - (l + 1/2)^2 R^{-2}$):
\EQ{
	\frac{\tilde{\mu_1} - \tilde{\mu_0}}{2} \pm \sqrt{  R^{-2} \left( l + \frac{1}{2} \pm \frac{R \tilde{\mu_0} + R \tilde{\mu_1}}{2} \right)^2 + \varphi^\dagger \varphi}.
}
The graph with $p \rightarrow -p$ just exchanges the roles of $\tilde{\mu_0}$ and $\tilde{\mu_1}$, so that together these two graphs yield
\EQ{
	\sqrt{  R^{-2} \left( l + \frac{1}{2} \pm \frac{R \tilde{\mu_0} + R \tilde{\mu_1}}{2} \right)^2 + \varphi^\dagger \varphi} \pm \frac{\tilde{\mu_1} - \tilde{\mu_0}}{2},
}
which becomes
\EQ{
	\sqrt{  R^{-2} \left( l + \frac{1}{2} \pm \frac{R \mu_1}{2} \right)^2 + \varphi^\dagger \varphi} \pm \frac{\mu_2 + \mu_3}{2}.
}
The two remaining graphs exchange $\varphi \rightarrow \varphi_\beta$ (for diagonal VEVs $\varphi_\beta^\dagger \varphi_\beta = \varphi_\beta \varphi_\beta^\dagger $), $\tilde{\mu}_0 \rightarrow \tilde{\mu}_3$, $\tilde{\mu}_1 \rightarrow \tilde{\mu}_2$, thus yielding:
\EQ{
	\sqrt{  R^{-2} \left( l + \frac{1}{2} \pm \frac{R \mu_1}{2} \right)^2 + \varphi_\beta^\dagger \varphi_\beta} \pm \frac{\mu_2 - \mu_3}{2}.
}

\section{Energies For Two VEVs}
In this section we will work out what the energy levels are when we turn on two VEVs
\EQ{
	\left\langle \phi_1 \right\rangle = \lambda^{(1)} U_{(N)}, \ \left\langle \phi_2 \right\rangle = \lambda^{(2)} V_{(N)},
}
where $\lambda^{(1)}$, $\lambda^{(2)}$, and $\lambda^{(3)}$ are complex numbers, and
\AL{
	U_{(N)} &= \textnormal{diag} \left( \omega, \omega^2, \ldots, \omega^N \right) \\
	\left(V_{(N)}\right)_{ab} &=
	\begin{cases}
	1 \textnormal{	if } b=a+1 \textnormal{ mod } N, \\
	0 \textnormal{ otherwise}
	\end{cases},
}
with $\omega = e^{2\pi i \beta}$. In other words, we have put $n=N$. Defining
\EQ{
	J_{a,b} \equiv V^a U^{-b} \omega^{\frac{ab}{2}},
}
we have the following useful relations:
\AL{
	[J_{a,b},J_{c,d}] &= 2 \sin \left( \frac{(bc-ad)\pi}{n} \right) J_{a+c,b+d} \\
	[J_{a,b},J_{c,d}]_{\pm \beta} &= 2 \sin \left( \frac{(bc-ad \pm 1)\pi}{n} \right) J_{a+c,b+d} \\
	J_{a,b}^\dagger &= J_{-a,-b} \\
	\Tr \left( J_{a,b}^\dagger J_{c,d} \right) &= n \delta_{ac} \delta_{bd} \\
	U &= J_{0,-1} \\
	V &= J_{1,0}
}
We expand the fields as
\EQ{
	\phi = \frac{1}{\sqrt{2N}} \sum_{i,j=1}^N \phi_{i,j} J_{i,j},
}
where $\phi_{i,j}$ is complex. For hermitian $\phi$, we have
\EQ{
	\phi_{i,j}^\dagger = \phi_{-i,-j}.
}

\subsection{Scalars}

At second order, the contribution from the D-term comes from
\EQ{
	V_D^{(2)} = \frac{1}{g^2} \Tr \left( [\varphi_1^\dagger,\phi_1] - [\varphi_1,\phi_1^\dagger] + [\varphi_2^\dagger,\phi_2] - [\varphi_2,\phi_2^\dagger] \right)^2,
}
while the gauge fixing contributes
\EQ{
	V_{gf}^{(2)} = - \frac{1}{g^2} \Tr \left( [\varphi_1^\dagger,\phi_1] + [\varphi_1,\phi_1^\dagger] + [\varphi_2^\dagger,\phi_2] + [\varphi_2,\phi_2^\dagger] \right)^2.
}
Together, this becomes
\SP{
	V_D^{(2)} + V_{gf}^{(2)} = - \frac{4}{g^2} \Tr \bigg( [\varphi_1,\phi_1^\dagger] [\varphi_1^\dagger,\phi_1] + [\varphi_2,\phi_2^\dagger] [\varphi_2^\dagger,\phi_2] + \\ + [\varphi_1,\phi_1^\dagger] [\varphi_2^\dagger,\phi_2] + [\varphi_2,\phi_2^\dagger] [\varphi_1^\dagger,\phi_1] \bigg),
}
which in commutator action notation is equal to
\SP{
	V_D^{(2)} + V_{gf}^{(2)} =
	\frac{1}{g^2} 2 \Tr \bigg( \phi_1^\dagger ( 2 \varphi_1 \varphi_1^\dagger ) \phi_1 +
	\phi_2^\dagger ( 2 \varphi_2 \varphi_2^\dagger ) \phi_2 + \\ +
	\phi_1^\dagger ( 2 \varphi_1 \varphi_2^\dagger ) \phi_2 +
	\phi_2^\dagger ( 2 \varphi_2 \varphi_1^\dagger ) \phi_1 \bigg)
}
At second order, the superpotential contributes
\EQ{
	V_W = \frac{4}{g^2} \, \Tr \left( | [\varphi_1,\phi_2]_\beta + [\phi_1,\varphi_2]_\beta |^2 + | [\varphi_2,\phi_3]_\beta |^2 + | [\phi_3,\varphi_1]_\beta |^2 \right),
}
which in commutator action notation is
\SP{
	V_W^{(2)} = \frac{1}{g^2} 2 \Tr \bigg( \phi_1^\dagger (2 (\varphi_2)_{-\beta}^\dagger (\varphi_2)_{-\beta} ) \phi_1 + \phi_2^\dagger (2 (\varphi_1)_\beta^\dagger (\varphi_1)_\beta ) \phi_2 + \\
	+ \phi_2^\dagger (2 (\varphi_1)_\beta^\dagger (\varphi_2)_{-\beta} ) \phi_1 
	+ \phi_1^\dagger (2 (\varphi_2)_{-\beta}^\dagger (\varphi_1)_\beta ) \phi_2 + \\
	+ \phi_3^\dagger ( 2 (\varphi_1)_{-\beta}^\dagger (\varphi_1)_{-\beta} ) \phi_3 + \phi_3^\dagger ( 2 (\varphi_2)_\beta^\dagger (\varphi_2)_\beta ) \phi_3 \bigg).
}
We will now use the following relations:
\AL{
	&[J_{a,b},\phi] = \frac{1}{\sqrt{2N}} \sum_{i,j=1}^N 2 \sin \left( \frac{(bi-aj)\pi}{n} \right) \phi_{i,j} J_{a+i,b+j} \\
	&[J_{a,b},\phi]_{\pm \beta} = \frac{1}{\sqrt{2N}} \sum_{i,j=1}^N 2 \sin \left( \frac{(bi-aj \pm 1)\pi}{n} \right) \phi_{i,j} J_{a+i,b+j} \\
	&[J_{a,b},V^{k} \phi U^{l}] = \nonumber \\ &= \frac{1}{\sqrt{2N}} \sum_{i,j=1}^N 2 \sin \left( \frac{(b(i+k)-a(j-l))\pi}{n} \right) \phi_{i,j} J_{a+i+k,b+j-l} \\
	&[J_{a,b},V^{k} \phi U^{l}]_{\pm \beta} = \nonumber \\ &= \frac{1}{\sqrt{2N}} \sum_{i,j=1}^N 2 \sin \left( \frac{(b(i+k)-a(j-l) \pm 1)\pi}{n} \right) \phi_{i,j} J_{a+i+k,b+j-l}
}
After a change of basis
\AL{
	\phi_1' \equiv \phi_1 U^\dagger, \\
	\phi_2' \equiv V^\dagger \phi_2 \\
	\phi_3' \equiv V \phi_3 U,
}
matters simplify, and we will see that the cross terms between $\phi_1$ and $\phi_2$ cancel. We have that
\SP{
	(\varphi_2)_{-\beta} \phi_1 &= \lambda^{(2)} [J_{1,0}, \phi_1' U]_{-\beta} = \\ &= - \frac{1}{\sqrt{2N}} \sum_{i,j=1}^N 2\lambda^{(2)} \sin \left( \frac{j\pi}{n} \right) \phi'^{(1)}_{i,j} J_{i+1,j-1}
}
\SP{
	(\varphi_1)_\beta \phi_2 &= \lambda^{(1)} [J_{0,-1}, V \phi_1']_\beta = \\ &= - \frac{1}{\sqrt{2N}} \sum_{i,j=1}^N 2\lambda^{(1)} \sin \left( \frac{i\pi}{n} \right) \phi'^{(2)}_{i,j} J_{i+1,j-1}
}
\SP{
	\varphi_1^\dagger \phi_1 &= \lambda^{(1)} [J_{0,1}, \phi_1' U] = \\ &= \frac{1}{\sqrt{2N}} \sum_{i,j=1}^N 2\lambda^{(1)} \sin \left( \frac{i\pi}{n} \right) \phi'^{(1)}_{i,j} J_{i,j}
}
\SP{
	\varphi_2^\dagger \phi_2 &= \lambda^{(2)} [J_{-1,0}, V \phi_1'] = \\ &= \frac{1}{\sqrt{2N}} \sum_{i,j=1}^N 2\lambda^{(1)} \sin \left( \frac{j\pi}{n} \right) \phi'^{(2)}_{i,j} J_{i,j}
}
\SP{
	(\varphi_1)_{-\beta} \phi_3 &= \lambda^{(1)} [J_{0,-1}, V^\dagger \phi_2 U^\dagger]_{-\beta} = \\ &= - \frac{1}{\sqrt{2N}} \sum_{i,j=1}^N 2 \lambda^{(1)} \sin \left( \frac{i\pi}{n} \right) \phi'^{(3)}_{i,j} J_{i-1,j}
}
\SP{
	(\varphi_2)_\beta \phi_3 &= \lambda^{(2)} [J_{1,0}, V^\dagger \phi_2 U^\dagger]_\beta = \\ &= - \frac{1}{\sqrt{2N}} \sum_{i,j=1}^N 2 \lambda^{(2)} \sin \left( \frac{j\pi}{n} \right) \phi'^{(3)}_{i,j} J_{i,j+1}
}
After rescaling $\lambda^{(1)} \rightarrow \lambda^{(1)} / \sqrt{2}$, $\lambda^{(2)} \rightarrow \lambda^{(2)} / \sqrt{2}$ and writing $\phi' \rightarrow \phi$, we finally get
\EQ{
	V^{(2)}_b = \sum_{a=1}^3 \sum_{i,j=1}^N \left( 4 \left[ |\lambda^{(1)}|^2 \sin^2 \left( \frac{i\pi}{n} \right) + |\lambda^{(2)}|^2 \sin^2 \left( \frac{j\pi}{n} \right) \right] {\phi^{(a)}_{i,j}}^\dagger \phi^{(a)}_{i,j} \right)
}
The rest of the analysis is analogous to the case with one VEV. (Indeed, it is completely the same as for $\mathcal{N}=4$ SYM.)

\subsection{Fermions}
The fermions get masses from coupling to gauginos
\EQ{
	\frac{1}{g^2} 2 \Tr \left( - i \sqrt{2} \lambda [\varphi_1^\dagger,\chi_1] - i \sqrt{2} \lambda [\varphi_2^\dagger,\chi_2] \right) + c.c.,
}
and from the superpotential
\EQ{
	\frac{1}{g^2} 2 \Tr \left( - i \sqrt{2} \chi_1 [\varphi_2,\chi_3]_\beta - i \sqrt{2} \chi_3 [\varphi_1,\chi_2]_\beta  \right) + c.c.
}
We make the following change of basis:
\AL{
	\chi_1' \equiv \chi_1 U^\dagger \\
	\chi_2' \equiv V^\dagger \chi_2 \\
	\chi_3' \equiv V \chi_3 U.
}
Then, we have that
\EQ{
	[\varphi_1^\dagger,\chi_1] = \lambda^{(1)} [J_{0,-1}, \chi_1 U] = \frac{1}{\sqrt{2N}} \sum_{i,j=1}^N 2 \sin \left( \frac{i\pi}{n} \right) \chi^{(1)}_{i,j} J_{i,j}
}
\EQ{
	[\varphi_2^\dagger,\chi_2] = \lambda^{(2)} [J_{1,0}, V \chi_2] = \frac{1}{\sqrt{2N}} \sum_{i,j=1}^N 2 \sin \left( \frac{j\pi}{n} \right) \chi^{(2)}_{i,j} J_{i,j}
}
\EQ{
	[\varphi_2,\chi_3]_\beta = \lambda^{(2)} [J_{1,0}, V^\dagger \chi_3 U^\dagger]_\beta = - \frac{1}{\sqrt{2N}} \sum_{i,j=1}^N 2 \sin \left( \frac{j\pi}{n} \right) \chi^{(3)}_{i,j} J_{i,j+1}
}
\EQ{
	[\varphi_1,\chi_2]_\beta = \lambda^{(1)} [J_{0,-1}, V \chi_2]_\beta = - \frac{1}{\sqrt{2N}} \sum_{i,j=1}^N 2 \sin \left( \frac{i\pi}{n} \right) \chi^{(2)}_{i,j} J_{i+1,j-1}
}
As can be seen, everything is the same as $\mathcal{N}=4$ SYM. Therefore we can argue that because of the $SU(2) \times U(1)$ symmetry that remains unbroken by turning on two chemical potentials, everything works out exactly as in the case with one VEV. Similar calculations give the expected masses for the gauge bosons.

\section{TsT-Transformation}
The general rules for T-duality transformations are given in \cite{Bergshoeff:1995as}. We also found \cite{Frolov:2005dj} a useful reference for how to derive the action of a TsT-transformation on $g$ and $b$. T acts on $g$, $b$, and the dilaton $\phi$ as follows ($i,j > 1$):
\AL{
	g_{11} &\rightarrow \frac{1}{g_{11}} \\
	g_{ij} &\rightarrow g_{ij} - \frac{g_{1i} g_{1j} - b_{1i} b_{1j}}{g_{11}} \\
	g_{1i} &\rightarrow \frac{b_{1i}}{g_{11}} \\
	b_{ij} &\rightarrow b_{ij} - \frac{b_{1i} b_{1j} - b_{1i} g_{1j}}{g_{11}} \\
	b_{1i} &\rightarrow \frac{g_{1i}}{g_{11}} \\
	e^{2\phi} &\rightarrow \frac{e^{2\phi}}{g_{11}}
}
The shift s, given by
\EQ{
	\varphi_2 \rightarrow \varphi_2 + \gamma \varphi_1,
}
acts on $g$ as
\AL{
	g_{11} &\rightarrow g_{11} + \gamma^2 g_{22} + 2\gamma g_{12} \\
	g_{1i} &\rightarrow g_{1i} + \gamma g_{2i},
}
and on $b$ as
\EQ{
	b_{1i} \rightarrow b_{1i} + \gamma b_{2i}.
}
Starting with $b = 0$, a TsT-transformation gives for $i,j > 2$ ($G_{ij}$, $B_{ij}$ are the TsT-transformed fields):
\SP{
	G_{ij} = G g_{ij} + G \gamma^2 \bigg[ g_{ij}g_{22}g_{11} + g_{1i}g_{2j}g_{12} + g_{1j}g_{2i}g_{12} - \\ - g_{1i}g_{1j}g_{22} - g_{ij}g_{12}g_{12} - g_{2i}g_{2j}g_{11} \bigg],
}
where
\EQ{
	G \equiv \frac{1}{1 + \gamma^2 (g_{22}g_{11}-g_{12}^2)}.
}
For $i \leq 2$ or $j \leq 2$, we have
\EQ{
	G_{ij} = G g_{ij}.
}
For $b$, a TsT-transformation gives
\EQ{
	B_{ij} = G \gamma ( g_{1i}g_{2j} - g_{1j}g_{2i} ).
}
(Note that if $g_{1i}=g_{2i}=g_{1j}=g_{2j}=0$, then $G_{ij} = g_{ij}$ and $B_{ij} = b_{ij}$.) The dilaton transforms as
\EQ{
	e^{2\phi} \rightarrow G e^{2\phi},
}
and, finally, the $n$-forms transform as \cite{Imeroni:2008cr}
\EQ{
    \sum_q C_q \wedge e^{-B} = \sum_q c_q \wedge e^{-b} + \gamma \left[ \sum_q c_q \wedge e^{-b}
\right]_{[\varphi^1][\varphi^2]},
}
where for a general $p$-form $\omega_p$ we have defined
\EQ{
	\omega_p = \bar \omega_p + \omega_{p[y]} \wedge dy,
}
where $\bar \omega_p$ does not contain any legs in $dy$.

\end{document}